\documentclass[french,english]{article}
\usepackage[T1]{fontenc}
\usepackage[latin9]{inputenc}
\usepackage{geometry}
\usepackage{array}
\usepackage{mathtools}
\usepackage{amsmath}
\usepackage{amssymb}

\makeatletter

\providecommand{\tabularnewline}{\\}

\newcommand{\lyxaddress}[1]{
\par {\raggedright #1
\vspace{1.4em}
\noindent\par}
}
\newenvironment{lyxlist}[1]
{\begin{list}{}
{\settowidth{\labelwidth}{#1}
 \setlength{\leftmargin}{\labelwidth}
 \addtolength{\leftmargin}{\labelsep}
 }}
{\end{list}}

\usepackage{babel}

\usepackage{babel}

\addto\extrasfrench{%
   \providecommand{\fg}{\ifdim\lastskip>\z@\unskip\fi~\frqq}%
}

\makeatother

\usepackage{babel}
\makeatletter
\addto\extrasfrench{%
   \providecommand{\fg}{\ifdim\lastskip>\z@\unskip\fi~\frqq}%
}

\makeatother
\begin{document}

\title{Eigenlogic \\
 in the spirit of George Boole}

\author{Zeno TOFFANO}
\maketitle

\lyxaddress{\begin{center}
\textit{CentraleSupelec - Laboratoire des Signaux et Systèmes (L2S-UMR8506)
- CNRS - Université Paris-Saclay}\\
 \textit{3 rue Joliot-Curie, F-91190 Gif-sur-Yvette, FRANCE \ \ \ \ \ \ \ \ \ \ \ \ \ \ \ zeno.toffano@centralesupelec.fr} 
\par\end{center}}
\begin{abstract}
This work presents an operational and geometric approach to logic.
It starts from the multilinear elective decomposition of binary logical
functions in the original form introduced by George Boole. A justification
on historical grounds is presented bridging Boole's theory and the
use of his arithmetical logical functions with the axioms of Boolean
algebra using sets and quantum logic. It is shown that this algebraic
polynomial formulation can be naturally extended to operators in finite
vector spaces. Logical operators will appear as commuting projection
operators and the truth values, which take the binary values $\{0,1\}$,
are the respective eigenvalues. In this view the solution of a logical
proposition resulting from the operation on a combination of arguments
will appear as a selection where the outcome can only be one of the
eigenvalues. In this way propositional logic can be formalized in
linear algebra by using elective developments which correspond here
to combinations of tensored elementary projection operators. The original
and principal motivation of this work is for applications in the new
field of quantum information, differences are outlined with more traditional
quantum logic approaches. 
\end{abstract}

\section{Introduction}

The year 2015 celebrated discretely the 200\textsuperscript{th} anniversary
of the birth of George Boole (1815-1864). His visionary approach to
logic has led to the formalization in simple mathematical language
what was prior to him a language and philosophy oriented discipline.
His initial motivation as it appears clearly in his first work on
logic in 1847: \textit{Mathematical Analysis of Logic} \cite{key-1},
was to propose an algebraic formulation which could generate all the
possible logical propositions, to express any logical proposition
by an equation, and find the most general consequences of any finite
collection of logical propositions by algebraic reasoning applied
to the corresponding equations. He then wrote the synthesis of all
his investigations in logic in 1854 with the \textit{The Laws of Thought}
\cite{key-2}.

In 1847 George Boole was already an outstanding mathematician, he
was awarded the Gold Medal of the Royal Society in 1844 for his memoir
\textit{On a General Method in Analysis}. He was an expert in the
resolution of nonlinear differential equations and introduced many
new methods using symbolic algebra as stated by Maria Panteki \cite{key-3}.
Evidently George Boole became fond of operators because of his successes
in applying the algebra of differential operators in the years 1841\textemdash 1845.

His approach can be viewed as operational, this characteristic is
rarely considered nowadays as pointed out by Theodeore Halperin \cite{key-4,key-5}.
George Boole (see \cite{key-1} p.$16$) uses $X$, $Y$, $Z$...
to represent the individual members of classes. He then introduces
the symbol $x$, which he named \textit{elective symbol}, operating
upon any object comprehending individuals or classes by selecting
all the $X$'s which it contains. It follows that the product of the
elective symbols ``$xy$ will represent, in succession, the selection
of the class $Y$, and the selection from the class $Y$ of such objects
of the class $X$ that are contained in it, the result being the class
common to both $X$'s and $Y$'s''. In logical language this is the
operation of conjunction, $AND$.

An expression in which the elective symbols, $x$, $y$, $z$...,
are involved becomes an elective function only if it can be considered
``interpretable'' in logic. George Boole did not give a precise
definition of what he meant by an elective function, it seems likely
that he meant that any algebraic function in elective symbols $x$,
$y$, $z$..., would be an elective function. This is the case when
the expression sums up to the two possible values $0$ and $1$. In
logic the numbers $0$ and $1$ correspond to false and true respectively.
So, according to George Boole, all the quantities become interpretable
when they take the values $0$ and $1$.

George Boole's logic using symbolic algebra was different and new
because he was convinced that logic had not only to do with ``quantity''
but should possess a ``deeper system of relations'' that had to
do with the activity of ``deductive reasoning''. Now with these
premises he was able to use all the common operations of ordinary
algebra but introducing a special condition on the symbols : the idempotence
law. This law can only be satisfied by the numbers $0$ and $1$ and
was by him considered as the peculiar law for logic. In his second
book on logic \cite{key-2} he gives to this law the status of ``the
fundamental law of thought''.

For George Boole all arguments and functions in logic can be considered
as elective symbols. For example he stated (p.$63$ in \cite{key-1})
: ``It is evident that if the number of elective symbols is $n$,
the number of the moduli will be $2^{n}$, and that their separate
values will be obtained by interchanging in every possible way the
values $1$ and $0$ in the places of the elective symbols of the
given function.'' $n$ stands for the number of elective symbols
which correspond to the number of arguments of the logical system
or in modern language its \textit{arity} (the letter $m$ in the original
text is here replaced by the letter $n$). The \textit{moduli}, for
George Boole, are the co-factors in the development (p. 62 in \cite{key-1}).
From what stated above an obvious conclusion is that there are $2^{2^{n}}$possible
expansions of elective functions, but curiously George Boole does
not draw this conclusion explicitly.

With the introduction of truth tables by Charles Sanders Peirce in
the early 1880's \cite{key-6}, attracting little attention at the
time, as stated by Karl Menger in \cite{key-8-1} and successively,
around 1920, rediscovered simultaneously and independently by Emil
Post \cite{key-7} and by Ludwig Wittgenstein (5.101 in \textit{Tractatus}
\cite{key-8}) the counting of the number of possible elementary logical
propositions (connectives) became an evidence. By the way Emil Post
in \cite{key-7} extended the counting to alphabets greater than binary
($m>2$) leading to the combinatorial number $m^{m^{n}}$ of elementary
multi-valued logical connectives with $m$ values and $n$ arguments.

The aspect of Boole's method which has been much discussed was his
interpretation given to the two special numbers: $1$ and $0$. The
number $1$ represented for him the class of all conceivable objects
i.e. the entire universe, and naturally the number $0$ should have
represented the empty class. But it is not clear in \cite{key-1,key-2}
if George Boole does ever refer to $0$ as being a class, or was it
just part of his algebraic machinery? As for the objection to the
use of $1$, it has been to the requirement that it refer to the entire
universe as opposed to a \textit{universe of discourse} (extent of
the field within which all the objects of our discourse are found)
\cite{key-41}.

George Boole introduces considerable vagueness in \cite{key-1} as
to when one is working in a logic of classes, and when in a logic
of propositions. In his propositional calculus he restricted his attention
to statements that were always true or always false, this reduces
hypothetical propositions to categorical propositions. In 1854 \cite{key-2}
George Boole more explicitly replaces the algebra of selection operators
by the algebra of classes.

In this paper hypothetical propositions will not be considered, the
analysis will be restricted to what is currently named \textit{propositional
logic} (also named \textit{sentential logic}) and will not deal with
\textit{predicate logic} (also named \textit{first-order logic}) which
uses \textit{quantifiers} (the \textit{existence quantifier} $\exists$
and the \textit{universal quantifier} $\forall$) on propositions.

As was outlined by Theodore Hailperin in \cite{key-4} the elective
symbols and functions denote operators and it will be emphasized in
this work that the algebra of elective symbols can also be interpreted
as an algebra of commuting projection operators and used for developing
propositional logic in a linear algebra framework by the isomorphism
of Boole's elective symbols and functions with commuting projection
operators.

\section{\label{sec:Elective-functions}Elective symbols and functions}

\subsection{Idempotence and Boole's development theorem}

Here are briefly presented the basic concepts underlying the elective
decomposition method, starting from the very first intuition of George
Boole regarding his digitization of logic.

Elective symbols obey the following laws, these are sufficient to
build an algebra.

Law (\ref{eq:Distributiv}) says that elective symbols are distributive.
This means, according to Boole, that ``the result of an act of election
is independent of the grouping or classification of the subject''.

\begin{equation}
x(u+v)=xu+xv\label{eq:Distributiv}
\end{equation}

Law (\ref{eq:Commutativ}) says that elective symbols commute, this
because: ``it is indifferent in what order two successive acts of
election are performed''.

\begin{equation}
xy=yx\label{eq:Commutativ}
\end{equation}

Law (\ref{eq:IndexLaw}) called\textit{ index law} by George Boole
represents the idempotence of an elective symbol, he states: ``that
the result of a given act of election performed twice or any number
of times in succession is the result of the same act performed once''.

\begin{equation}
x^{n}=x\label{eq:IndexLaw}
\end{equation}

As a consequence of this law George Boole formulated the two following
equivalent equations.

\begin{eqnarray}
x^{2} & = & x\nonumber \\
x(1-x) & = & 0\label{eq:idempotence}
\end{eqnarray}

Equation (\ref{eq:idempotence}) explicitly shows that the numbers
$0$ and $1$ are the only possible ones. It also states the orthogonality
between the elective symbol $x$ and $(1-x)$, which represents the
complement or negation of $x$. Also:

\begin{equation}
x+(1-x)=1\label{eq:complement}
\end{equation}
this equation shows that the symbol $x$ and its complement $(1-x)$
form the universe class.

Now with these laws and symbols elective functions can be calculated.
It is interesting to illustrate how George Boole came to a general
expression of an elective function using the Mac Laurin development
of the function $f(x)$ around the number $0$ (see \cite{key-1}
p.60). Because of the index law (\ref{eq:IndexLaw}) or the idempotence
law (\ref{eq:idempotence}) the symbol $x$ becomes a factor of the
series starting from the second term in the Mac Laurin development,
this gives:

\begin{equation}
f(x)=f(0)+x[f'(0)+\dfrac{1}{2!}f''(0)+\dfrac{1}{3!}f'''(0)+...]\label{eq:MacLaurin1}
\end{equation}

Then by calculating the function at the value $1$, $f(x=1)$, using
equation (\ref{eq:MacLaurin1}), one finds a substitute expression
of the series. By substituting this expression back in equation (\ref{eq:MacLaurin1})
one finally gets:

\begin{eqnarray}
f(x) & = & f(0)+x(f(1)-f(0))\nonumber \\
 & = & f(0)(1-x)+f(1)x\label{eq:ElectivDev1}
\end{eqnarray}

In a simpler way these expressions can be obtained directly by classical
interpolation methods using for example Lagrange interpolation polynomials
for a finite number $m$ of distinct points $x_{i}$. The Lagrange
polynomials being then of degree $m-1$ and are given by:

\begin{equation}
\pi_{x_{i}}(x)=\prod_{j\,(j\neq i)}^{m}\frac{(x-x_{j})}{(x_{i}-x_{j})}\label{eq:Lagr_polynom}
\end{equation}

The interpolation function $f(x)$ of a given function $g(x)$ is
then expressed using the finite polynomial development over the chosen
$m$ distinct points $x_{i}$ : 
\begin{equation}
f(x)=\sum_{i=1}^{m}g(x_{i})\,\pi_{x_{i}}(x)\label{eq:Lagr_interpol}
\end{equation}

For a binary system ($m=2$) with alphabet values $\{0,1\}$ the two
interpolation polynomials are easily calculated from (\ref{eq:Lagr_polynom}),
giving respectively: $\pi_{x_{0}=0}(x)=(1-x)$ and $\pi_{x_{1}=1}(x)=x$,
which are the same as in (\ref{eq:ElectivDev1}), and for this alphabet,
equation (\ref{eq:Lagr_interpol}) is equivalent to equation (\ref{eq:ElectivDev1})
because of course at the interpolation points $g(0)=f(0)$ and $g(1)=f(1)$.
In this way the demonstration of the elective development theorem
does not necessitate infinite polynomial power series, \textit{e.g.}
the Maclaurin expansion, as was done with the power series proof by
George Boole in \cite{key-1}.

It must be underlined that Lagrange polynomials (\ref{eq:Lagr_polynom})
are by construction idempotent functions at the interpolation points,
more precisely: $\pi_{x_{i}}(x=x_{i})=1$ and $\pi_{x_{i}}(x=x_{j}\neq x_{i})=0$.
The same interpolation method can be extended to other binary alphabets,
e.g. $\{+1,-1\}$, and also for multi-valued systems with $m>2$ (for
developments see \cite{key-9}).

Equation (\ref{eq:ElectivDev1}) shows that an elective function can
be uniquely developed using the two orthogonal elective symbols $x$
and $(1-x)$. Now if the function is to be ``interpretable'' in
logic it should only take the values $0$ and $1$, and this means
that both co-factors $f(0)$ and $f(1)$ (\textit{moduli} for George
Boole) take also the values $0$ or $1$. These coefficients represent
the \textit{truth values} for the logical function.

How many possibilities, or stated in logical language, how many different
logical functions can we build using $n$ arguments? We already discussed
that the possible combinations are $2^{2^{n}}$. So considering a
unique symbol, $n=1$, one obtains $4$ distinct elective functions.
These are shown on table \ref{Table 1}.

A similar procedure can be used (see p.$62$ in \cite{key-1}) for
elective functions of two arguments $f(x,y)$, this gives the following
multilinear development using $4$ orthogonal and idempotent polynomials:

\begin{equation}
f(x,y)=f(0,0)\,(1-x)(1-y)+f(0,1)\,(1-x)y+f(1,0)\,x(1-y)+f(1,1)\,xy\label{eq:ElectDev2}
\end{equation}

And so on for increasing $n$. For $n=2$ one has $2^{2^{n=2}}=16$
different elective functions (given in Table \ref{Table 2}) and for
$n=3$, $2^{2^{n=3}}=256$. All elective functions are idempotent:
$f_{el}^{2}=f_{el}$. Here also finite interpolation methods could
be used this time using multivariate functions.

Equation (\ref{eq:ElectDev2}) represents the canonical elective development
of a two argument elective function and has the same structure as
the \textit{minterm} disjunction canonical form in Boolean algebra
\cite{key-4} which represents the disjunction of mutually exclusive
conjunctions (see hereafter).

So from equation (\ref{eq:ElectDev2}) all logical functions can be
expressed as a combination of degree $1$ multilinear polynomials.
It can be shown that this decomposition is unique.

George Boole has also developed a method of resolution of what he
called \textit{elective equations} where for example the question
is: for what values an elective function is true? (see \cite{key-1}
p. 70).

A very simple method used for resolving elective equations uses the
orthogonality of the different elective polynomials which are multiplied
by the respective co-factors (\textit{moduli}) $f^{[n]}(a,b,c,...)$
in the development, these polynomials are named $\pi_{(a,b,c...)}^{[n]}$
for a given combination of fixed values $(a,b,c,...)$. This gives
the following equation for selecting the individual co-factors for
an $n$ symbol elective function:

\begin{equation}
f^{[n]}(x,y,z,...)\cdot\pi_{(a,b,c,...)}^{[n]}=f^{[n]}(a,b,c,...)\:\pi_{(a,b,c,...)}^{[n]}\label{eq:ElEqProj}
\end{equation}

Equation (\ref{eq:ElEqProj}) can be used whatever the number of symbols
and also when the functions are not explicitly put in the canonical
form. For example if one wants to select the coefficient $f(0,1)$
out of $f(x,y)$ in equation (\ref{eq:ElectDev2}), one simply multiplies
the function by the corresponding orthogonal polynomial $(1-x)y$.
Without doubt it is most of the times easier to evaluate directly
$f(0,1)$.

\section{\label{sec:Elective-symbolic-logic}Elective symbolic logic}

\subsection{Truth tables and elective functions}

In this section the link of elective functions with ordinary propositional
logic is presented. Functions and symbols will take exclusively the
two binary values $0$ and $1$ representing respectively the false
($F$) and true ($T$) character of a given proposition. Logical functions
are classified according to their truth tables.

Starting from the very simple propositions derived from the single
elective symbol $x$, according to the function development in equation
(\ref{eq:ElectivDev1}), one sees that there are $4$ possible functions
depending on the values taken by $f(0)$ and $f(1)$ respectively.
This is shown on table \ref{Table 1}:

\begin{table}
\begin{centering}
\begin{tabular}{|>{\centering}p{1.3cm}||c||>{\centering}p{2cm}||>{\centering}p{2.5cm}||>{\centering}p{2cm}|}
\hline 
{\small{}func}t. {\footnotesize{}$f_{i}^{[1]}$}  & {\small{}logical proposition}  & {\small{}truth table }\foreignlanguage{french}{{\small{}$f(0)\;f(1)$}}  & {\small{}canonical form $(1-x)\;,\;x$ }  & {\small{}polynomial form}\tabularnewline
\hline 
\hline 
{\footnotesize{}$f_{0}^{[1]}$}  & {\footnotesize{}$F$}  & \foreignlanguage{french}{{\footnotesize{}$0\quad0$}\foreignlanguage{english}{  }} & {\footnotesize{}$0$}  & {\footnotesize{}$0$}\tabularnewline
\hline 
\hline 
{\footnotesize{}$f_{1}^{[1]}$}  & {\footnotesize{}$\bar{A}$}  & \foreignlanguage{french}{{\footnotesize{}$1\quad0$}\foreignlanguage{english}{  }} & {\footnotesize{}$(1-x)$}  & {\footnotesize{}$1-x$}\tabularnewline
\hline 
\hline 
{\footnotesize{}$f_{2}^{[1]}$}  & {\footnotesize{}$A$}  & \foreignlanguage{french}{{\footnotesize{}$0\quad1$}\foreignlanguage{english}{  }} & {\footnotesize{}$x$}  & {\footnotesize{}$x$}\tabularnewline
\hline 
\hline 
{\footnotesize{}$f_{3}^{[1]}$}  & {\footnotesize{}$T$}  & \foreignlanguage{french}{{\footnotesize{}$1\quad1$}\foreignlanguage{english}{  }} & {\footnotesize{}$(1-x)+x$}  & {\footnotesize{}$1$}\tabularnewline
\hline 
\end{tabular}
\par\end{centering}
{\small{}\caption{The four single argument logical elective functions}
\label{Table 1}}{\small \par}

\end{table}

In this case the two non trivial propositions are the \textit{logical
projection} $A$ and its negation $\bar{A}$. The other two give constant
outcomes: false $F$ and true $T$ whatever the value of the argument.

On Table \ref{Table 2} are shown the $16$ elective functions, $f_{i}^{[2]}$
, for $n=2$ arguments. The corresponding elective polynomials can
be straightforwardly obtained by substituting the respective truth
values in front of the four polynomial terms in equation (\ref{eq:ElectDev2}).
According to the standard classification, given for example by Donald
Knuth \cite{key-10}, logical functions are ordered with increasing
binary number in the truth table (counting order goes from left to
right: the lower digit is on the left). The representation used here
corresponds to what is often called the \textit{truth vector} of the
function: $(f(0,0),f(0,1),f(1,0),f(1,1))$.

$f_{0}^{[2]}$ has the truth values $(0,0,0,0)$ and represents contradiction,
$f_{1}^{[2]}$ is $NOR$ with truth values $(1,0,0,0)$ and so on...
For example conjunction ($AND$, \foreignlanguage{french}{{\small{}$\land$}})
is $f_{8}^{[2]}$ with $(0,0,0,1)$, disjunction ($OR$, \foreignlanguage{french}{{\small{}$\vee$}})
is $f_{14}^{[2]}$ with $(0,1,1,1)$ and exclusive disjunction ($XOR$,
\foreignlanguage{french}{{\small{}$\oplus$}}) is $f_{6}^{[2]}$ with
$(0,1,1,0)$.

In table \ref{Table 2} are also shown the canonical polynomial forms
issued directly from eq. (\ref{eq:ElectDev2}) and the respective
simplified polynomial expressions.

Some precisions on other logical connectives: the expression $A\Rightarrow B$
signifies ``$A$ implies $B$'', and the converse $A\Leftarrow B$
signifies ``$B$ implies $A$'' the symbol $\nRightarrow$ signifies
non-implication. The expression for $NAND$ which is ``not $AND$''
is given according to the De Morgan's law \cite{key-10} by $\bar{A}\lor\bar{B}$
. The same for $NOR$, ``not $OR$'', given by $\bar{A}\land\bar{B}$.

\medskip{}

\begin{table}
\begin{centering}
\begin{tabular}{|>{\centering}b{1cm}||>{\centering}b{2.8cm}||>{\centering}b{3.5cm}||>{\centering}b{5cm}||>{\centering}b{2cm}|}
\hline 
{\small{}funct. $f_{i}^{[2]}$}  & {\small{}logical connective for $A$ and $B$}  & {\small{}truth table}{\footnotesize{} }\foreignlanguage{french}{{\footnotesize{}$f(0,0)\;f(0,1)\;f(1,0)\;f(1,1)$}}  & {\small{}canonical form $(1-x)(1-y)\:,\:(1-x)y\:,\:x(1-y)\:,\:xy$}  & {\small{}polynomial form}\tabularnewline
\hline 
\hline 
{\small{}$f_{0}^{[2]}$}  & {\small{}$F$}  & \foreignlanguage{french}{{\footnotesize{}$0\;\;\;\;\;0\;\;\;\;\;0\;\;\;\;\;0$}\foreignlanguage{english}{
 }} & {\footnotesize{}$0$}  & {\footnotesize{}$0$}\tabularnewline
\hline 
\hline 
{\small{}$f_{1}^{[2]}$}  & {\small{}$NOR\;,\;\bar{A}\land\bar{B}$}  & \foreignlanguage{french}{{\footnotesize{}$1\;\;\;\;\;0\;\;\;\;\;0\;\;\;\;\;0$}\foreignlanguage{english}{
 }} & {\footnotesize{}$(1-x)(1-y)$}  & {\footnotesize{}$1-x-y+xy$}\tabularnewline
\hline 
\hline 
{\small{}$f_{2}^{[2]}$}  & {\small{}$A\nLeftarrow B$}  & \foreignlanguage{french}{{\footnotesize{}$0\;\;\;\;\;1\;\;\;\;\;0\;\;\;\;\;0$}\foreignlanguage{english}{
 }} & {\footnotesize{}$(1-x)y$}  & {\footnotesize{}$y-xy$}\tabularnewline
\hline 
\hline 
{\small{}$f_{3}^{[2]}$}  & {\small{}$\bar{A}$}  & \foreignlanguage{french}{{\footnotesize{}$1\;\;\;\;\;1\;\;\;\;\;0\;\;\;\;\;0$}\foreignlanguage{english}{
 }} & {\footnotesize{}$(1-x)(1-y)+(1-x)y$}  & {\footnotesize{}$1-x$}\tabularnewline
\hline 
\hline 
{\small{}$f_{4}^{[2]}$}  & {\small{}$A\nRightarrow B$}  & \foreignlanguage{french}{{\footnotesize{}$0\;\;\;\;\;0\;\;\;\;\;1\;\;\;\;\;0$}\foreignlanguage{english}{
 }} & {\footnotesize{}$x(1-y)$}  & {\footnotesize{}$x-xy$}\tabularnewline
\hline 
\hline 
{\small{}$f_{5}^{[2]}$}  & {\small{}$\bar{B}$}  & \foreignlanguage{french}{{\footnotesize{}$1\;\;\;\;\;0\;\;\;\;\;1\;\;\;\;\;0$}\foreignlanguage{english}{
 }} & {\footnotesize{}$(1-x)(1-y)+x(1-y)$}  & {\footnotesize{}$1-y$}\tabularnewline
\hline 
\hline 
{\small{}$f_{6}^{[2]}$}  & {\small{}$XOR\;,\;A\oplus B$}  & \foreignlanguage{french}{{\footnotesize{}$0\;\;\;\;\;1\;\;\;\;\;1\;\;\;\;\;0$}\foreignlanguage{english}{
 }} & {\footnotesize{}$(1-x)y+x(1-y)$}  & {\footnotesize{}$x+y-2xy$}\tabularnewline
\hline 
\hline 
{\small{}$f_{7}^{[2]}$}  & {\small{}$NAND\;,\;\bar{A}\vee\bar{B}$}  & \foreignlanguage{french}{{\footnotesize{}$1\;\;\;\;\;1\;\;\;\;\;1\;\;\;\;\;0$}\foreignlanguage{english}{
 }} & {\footnotesize{}$(1-x)(1-y)+(1-x)y+x(1-y)$}  & {\footnotesize{}$1-xy$}\tabularnewline
\hline 
\hline 
{\small{}$f_{8}^{[2]}$}  & {\small{}$AND\;,\;A\wedge B$}  & \foreignlanguage{french}{{\footnotesize{}$0\;\;\;\;\;0\;\;\;\;\;0\;\;\;\;\;1$}\foreignlanguage{english}{
 }} & {\footnotesize{}$xy$}  & {\footnotesize{}$xy$}\tabularnewline
\hline 
\hline 
{\small{}$f_{9}^{[2]}$}  & {\small{}$A\equiv B$}  & \foreignlanguage{french}{{\footnotesize{}$1\;\;\;\;\;0\;\;\;\;\;0\;\;\;\;\;1$}\foreignlanguage{english}{
 }} & {\footnotesize{}$(1-x)(1-y)+xy$}  & {\footnotesize{}$1-x-y+2xy$}\tabularnewline
\hline 
\hline 
{\small{}$f_{10}^{[2]}$}  & {\small{}$B$}  & \foreignlanguage{french}{{\footnotesize{}$0\;\;\;\;\;1\;\;\;\;\;0\;\;\;\;\;1$}\foreignlanguage{english}{
 }} & {\footnotesize{}$(1-x)y+xy$}  & {\footnotesize{}$y$}\tabularnewline
\hline 
\hline 
{\small{}$f_{11}^{[2]}$}  & {\small{}$A\Rightarrow B$}  & \foreignlanguage{french}{{\footnotesize{}$1\;\;\;\;\;1\;\;\;\;\;0\;\;\;\;\;1$}\foreignlanguage{english}{
 }} & {\footnotesize{}$(1-x)(1-y)+(1-x)y+xy$}  & {\footnotesize{}$1-x+xy$}\tabularnewline
\hline 
\hline 
{\small{}$f_{12}^{[2]}$}  & {\small{}$A$}  & \foreignlanguage{french}{{\footnotesize{}$0\;\;\;\;\;0\;\;\;\;\;1\;\;\;\;\;1$}\foreignlanguage{english}{
 }} & {\footnotesize{}$x(1-y)+xy$}  & {\footnotesize{}$x$}\tabularnewline
\hline 
\hline 
{\small{}$f_{13}^{[2]}$}  & {\small{}$A\Leftarrow B$}  & \foreignlanguage{french}{{\footnotesize{}$1\;\;\;\;\;0\;\;\;\;\;1\;\;\;\;\;1$}\foreignlanguage{english}{
 }} & {\footnotesize{}$(1-x)(1-y)+x(1-y)+xy$}  & {\footnotesize{}$1-y+xy$}\tabularnewline
\hline 
\hline 
{\small{}$f_{14}^{[2]}$}  & {\small{}$OR\;,\;A\vee B$}  & \foreignlanguage{french}{{\footnotesize{}$0\;\;\;\;\;1\;\;\;\;\;1\;\;\;\;\;1$}\foreignlanguage{english}{
 }} & {\footnotesize{}$(1-x)y+x(1-y)+xy$}  & {\footnotesize{}$x+y-xy$}\tabularnewline
\hline 
\hline 
{\small{}$f_{15}^{[2]}$}  & {\small{}$T$}  & \foreignlanguage{french}{{\footnotesize{}$1\;\;\;\;\;1\;\;\;\;\;1\;\;\;\;\;1$}\foreignlanguage{english}{
 }} & {\footnotesize{}$(1-x)(1-y)+(1-x)y+x(1-y)+xy$}  & {\footnotesize{}$1$}\tabularnewline
\hline 
\end{tabular}
\par\end{centering}
{\small{}\caption{The sixteen two argument logical elective functions}
\label{Table 2}}{\small \par}

\end{table}

Negation is obtained complementing the function by subtracting from
the number $1$.

The conjunction, $AND$, corresponds to the following elective function:

\begin{equation}
f_{8}^{[2]}(x,y)=f_{AND}^{[2]}(x,y)=xy
\end{equation}
{\small{} }and its negation $NAND$ is simply:

\begin{equation}
f_{7}^{[2]}(x,y)=1-xy=1-f_{AND}^{[2]}(x,y)=f_{NAND}^{[2]}(x,y)
\end{equation}

By complementing the input symbols i.e. by replacing the symbols $x$
and $y$ by $1-x$ and $1-y$ respectively one gets other logical
functions. For example considering:

\begin{eqnarray}
f_{1}^{[2]}(x,y) & = & (1-x)(1-y)=1-x-y-xy=1-(x+y-xy)\nonumber \\
 & = & 1-f_{14}^{[2]}(x,y)=1-f_{OR}^{[2]}(x,y)=f_{NOR}^{[2]}(x,y)
\end{eqnarray}
this is the complement of the disjunction $OR$ named $NOR$. This
result corresponds to De Morgan's law \cite{key-10} that states that
the conjunction $AND$ of the complements is the complement of the
disjunction $OR$.

\begin{equation}
f_{14}^{[2]}(x,y)=f_{OR}^{[2]}(x,y)=x+y-xy\label{eq:electiveOR}
\end{equation}
remark that the expression of the disjunction $OR$ is given by a
polynomial expression containing a minus sign, this is specific to
elective functions, and it must be this way in order that the functions
be ``interpretable''.

The expression for the exclusive disjunction $XOR$ is given by:

\begin{equation}
f_{6}^{[2]}(x,y)=f_{XOR}^{[2]}(x,y)=x+y-2xy\label{eq:electiveXOR}
\end{equation}
this form differs from what is usually used in logic where the last
term is omitted due to the fact that the addition operation is considered
a modulo $1$ sum in Boolean algebra. This function represents the
parity function giving $1$ when the total number of $1$'s of the
arguments is odd.

The function for implication (named \textit{material implication})
can also be obtained by the same method, the function corresponding
to $A\Rightarrow B$ will be $f_{\Rightarrow}^{[2]}$ and the converse
$f_{\Leftarrow}^{[2]}$. According to table \ref{Table 2}:

\begin{equation}
f_{\Rightarrow}^{[2]}(x,y)=f_{11}^{[2]}(x,y)=1-x+xy\qquad\qquad f_{\Leftarrow}^{[2]}(x,y)=f_{13}^{[2]}(x,y)=1-y+xy
\end{equation}
\\

Using De Morgan's theorem, by complementing the arguments, it is easy
to verify that $f_{\Rightarrow}^{[2]}$ transforms into $f_{\Leftarrow}^{[2]}$.

The non-implication cases will be respectively $f_{\nRightarrow}^{[2]}$
and $f_{\nLeftarrow}^{[2]}$ and are given by:

\begin{equation}
f_{\nRightarrow}^{[2]}(x,y)=f_{4}^{[2]}(x,y)=x-xy=1-f_{\Rightarrow}^{[2]}\qquad\qquad f_{\nLeftarrow}^{[2]}(x,y)=f_{2}^{[2]}(x,y)=y-xy=1-f_{\Leftarrow}^{[2]}
\end{equation}

One can of course go on by increasing the number of arguments $n$
in a straightforward way. Let's consider the cases $n=3$, the conjunction
becomes:

\begin{equation}
f_{AND}^{[3]}(x,y,z)=xyz\label{eq:AND3}
\end{equation}

The expression of disjunction is obtained in the same way as in equation
(\ref{eq:ElectDev2}) but with three elective symbols $x$,$y$ and
$z$. Doing straightforward calculation using the $8$ truth values
$(0,1,1,1,1,1,1,1)$ gives:

\begin{equation}
f_{OR}^{[3]}(x,y,z)=x+y+z-xy-xz-yz+xyz\label{eq:OR3 InclExcl}
\end{equation}
\\
 which represents the well-known inclusion-exclusion rule, and can
be extended to any arity $n$ by recurrence.

For the $XOR$ function with $n=3$ one gets, using the truth values
$(0,1,1,0,1,0,0,1)$:

\begin{equation}
f_{XOR}^{[3]}(x,y,z)=x+y+z-2xy-2xz-2yz+4xyz\label{eq:XOR3}
\end{equation}
this last expression represents a specific rule which can be extended
straightforwardly to any $n$ by recurrence.

Another very popular function for $n=3$ arguments is the majority
$MAJ$ which gives the value $1$ when there is a majority of $1$'s
for the arguments. The function is obtained using the truth values
$(0,0,0,1,0,1,1,1)$:

\begin{equation}
f_{MAJ}^{[3]}(x,y,z)=xy+xz+yz-2xyz\label{eq:MAJ3}
\end{equation}

These last two logical connectives are currently used together in
digital electronics to build a binary full-adder using logical gates,
the three input $XOR$ gives the binary sum and the three input $MAJ$
gives the carry out.

So it can be seen that this method is completely general and can be
straightforwardly applied to all logical connectives whatever the
number of arguments.

\subsection{Logical developments }

An idempotent elective function $f(x,y,...)$ can be evaluated at
the values $0$ and $1$ by using ordinary numerical algebra, and
all the usual propositional functions have truth tables that can be
expressed in either Boole's canonical form or polynomial form, \textit{e.g.},
one has $XOR\:\oplus$ expressed by $x(1-y)+(1-x)y$ as well as $x+y-2xy$.
An important remark must be made about the use of the two different
polynomial developments named respectively ``canonical form'' and
``polynomial form'' shown in the two last columns of Table \ref{Table 2}.
The canonical form corresponds to what is named in modern digital
logic the canonical minterm decomposition. The minterms correspond
here to products of elective polynomials. For example for $n=2$ arguments
the minterms are the $4$ orthogonal polynomials given in equation
(\ref{eq:ElectDev2}), in logical language each minterm is one of
the possible $4$ conjunctions obtained by compleme$R$ nting none,
one or two arguments.

One can always put whatever logical function in the canonical form
SOP (Sum Of Products), also named the \textit{full conjunctive normal
form} \cite{key-10} which is a sum of \textit{minterms}. A minterm
being formed by all input arguments, in a given combination complemented
or not, connected by conjunction $\wedge$ and ``Sum'' corresponding
to the disjunction $\vee$ (also exclusive disjunction $\oplus$,
as discussed hereafter). Another canonical decomposition is POS (Product
Of Sums) of \textit{maxterms}. A maxterm being all input arguments
connected by disjunction $\vee$, in a given combination complemented
or not, and ``Product'' corresponding to conjunction $\wedge$,
this form is also named the \textit{disjunctive normal form}.

A SOP with four input arguments can be considered for the following
working example:

\begin{eqnarray}
 & F_{\Sigma m(5,7,10,15)}^{[4]}(A,B,C,D)=\nonumber \\
 & (\overline{A}\wedge B\wedge\overline{C}\wedge D)\vee(\overline{A}\wedge B\wedge C\wedge D)\vee(A\wedge B\wedge\overline{C}\wedge D)\vee(A\wedge B\wedge C\wedge D)\label{eq:Minterm4-example}
\end{eqnarray}

The expression $\Sigma m(5,7,10,15)$ is the standard minterm notation,
where the numbers correspond to the specific minterms used in the
development. In this form one can easily verify that only one among
all minterms can be true at a time, this means that each disjunction
$\vee$ is actually an exclusive disjunction $\oplus$. In the minterm
SOP decomposition, because all the terms are orthogonal, disjunction
and exclusive disjunction play the same role.

One can write the expression given in equation (\ref{eq:Minterm4-example})
using the formalism presented in this paper by writing directly the
elective decomposition:

\begin{equation}
f_{\Sigma m(5,7,10,15)}^{[4]}(x,y,z,r)=(1-x)y(1-z)r+(1-x)yzr+xy(1-z)r+xyzr=yr
\end{equation}
so one can transform this expression into other polynomial forms in
order to get a simpler expression. Significant simplifications are
obtained when one can factor an argument and its complement for the
same expression, for example $x$ and $(1-x)$. The simplest case
being the logical projectors themselves such as $A$ in table \ref{Table 2}
where the canonical form $x(1-y)+xy$ reduces to $x$. This last argument
is essentially what is used to operate reduction of logical functions
by using Karnaugh maps \cite{key-10}.

\subsection{Discussion of elective arithmetic logic}

Characteristic of George Boole's method is that while some terms
appearing in logical expressions may be uninterpretable, equations
always are when suitably interpreted, by the rules$(+,-,\times,0,1)$,
leading \textit{in fine} to the values $0$ and $1$. He also recognizes
terms that cannot always be interpreted, such as the term $2xy$,
which arises in equation manipulations as for the elective function
corresponding to $XOR$ in (\ref{eq:electiveXOR}). The coherence
of the whole enterprise is justified in what Stanley Burris has later
called the \textquotedbl{}rule of $0$'s and $1$'s\textquotedbl{}
\cite{key-21}, which justifies the claim that uninterpretable terms
cannot be the ultimate result of equational manipulations from meaningful
starting formulae. George Boole provided no proof of this rule, but
the consistency of his system was later proved by Theodore Hailperin
\cite{key-4}, who provided an interpretation based on a fairly simple
construction of rings from the integers to provide an interpretation
of Boole's theory (see hereafter).

Even though this procedure is simple and straightforward it is not
in the habits of logic to use these arithmetic expressions, and the
reason why is not so clear. One explanation could be because of technology
driven habits: the development of computers using logical gates as
building blocks, and binary-digits (bits) as information units has
generalized what is called ``Boolean algebra'' formulated in its
actual form by Edward Huntington in 1904 \cite{key-11}, which is
not Boole's elective algebra \cite{key-5}. For example addition is
considered in Boolean algebra as a modulo $1$ sum giving: $x+x=x$.
For a Boolean ring we have even a different rule: $x+x=0$. Whereas
the elective calculation employs normal arithmetic addition and subtraction
as seen previously.

Arithmetic expressions are closely related to polynomial expressions
over the Galois field $GF_{2}=\mathbb{Z}/\mathbb{Z}_{2}$, but with
variables and function values interpreted as the integers $0$ and
$1$ instead of logic values. In this way, arithmetic expressions
can be considered as integer counterparts of polynomial expressions
over $GF_{2}$. For two Boolean variables $x_{1}$ and $x_{2}$ (using
here more standard notation corresponding to two bits) the necessary
relations are:

\begin{eqnarray}
 & \bar{x}=1-x & x_{1}\wedge x_{2}=x_{1}x_{2}\nonumber \\
 & x_{1}\vee x_{2}=x_{1}+x_{2}-x_{1}x_{2}\qquad & x_{1}\oplus x_{2}=x_{1}+x_{2}-2x_{1}x_{2}\label{eq:arithmetic-expr}
\end{eqnarray}
this resumes all the discussion of the preceding section, the right
part of the equations is called the \textit{arithmetic expression}.

It seems that, historically, only John Venn explicitly used the original
reasoning of George Boole in order to build his logical graphic diagrams
\cite{key-12}. He used surfaces on a $2$ dimensional space which
represented the different logical propositions and more precisely
intersection and union corresponding to conjunction and disjunction.
Doing this he had, in some cases, to subtract portions of surfaces
in order to get the correct surface measure. For example considering
two overlapping surfaces, the surface representing disjunction, $\vee$,
is obtained by the sum of the two surfaces minus their intersecting
surface (without this subtraction one would count twice the intersecting
surface), also for exclusive disjunction, $\oplus$, one has to subtract
twice the intersecting surface, this leads to formulae of the \textit{inclusion-exclusion}
type as illustrated in equations (\ref{eq:OR3 InclExcl}) and (\ref{eq:XOR3}).
The canonical forms of idempotent elective functions in Boole's algebra
are the same as for functions in Boolean algebra, and the number of
these were well-known in the second half of the 1800s, and fully written
out for three variables by John Venn in 1881 (according to Ernst Schroder
in \cite{key-31}).

In 1933 Hassler Whitney \cite{key-13}, showed how to convert the
modern algebra of classes (using union, intersection and complement)
into numerical algebra, giving three different normal forms (polynomials
in $x$'s, polynomials in $(1-x)$'s, and Boole's form) for functions.
He failed to recognize that he was converting the modern algebra of
classes into Boole's algebra of classes. Theodore Hailperin would
realize this decades later.

The observation that one can express propositional functions, viewed
as switching functions, using polynomials in ordinary numerical algebra,
as George Boole did, was used by Howard Aiken in 1951 in \cite{key-14},
where one finds tables for minimal ordinary numerical algebraic expressions
for switching functions $f\,:\,\left\{ 0,1\right\} ^{n}\,\rightarrow\,\left\{ 0,1\right\} $
up to $n=4$. It is interesting to note that Howard Aiken, who founded
the ``Harvard Computing Laboratory'', the first laboratory devoted
to Computer Science at Havard starting in 1937, developed the first
computer, the ASCC (Automatic Sequence Controlled Calculator), also
called Harvard MARK 1 in 1944 with IBM. He first found that arithmetic
expressions can be useful in designing logic circuits and used them
in the successive computers Harvard MARK 3 and MARK 4. This kind of
logic did not breakthrough principally because the family of Harvard
MARK computers where replaced by the ENIAC computer generation which
used semiconductor transistors instead of electromechanical switches
and vacuum tubes and relied on the \textit{bit} and \textit{logical
gate} paradigm introduced originally by Claude E. Shannon in 1938
\cite{key-51} where he ``tailored'' Boolean logic to switching
circuits. 

Nowadays these arithmetic developments are still used for describing
switching functions and decision logic design. A good review is given
by Svetlana Yanushkevich in \cite{key-15}. Arithmetic representations
of Boolean functions (\textit{i.e.} here elective functions) are known
as \textit{word-level forms}, and are a way to describe the parallel
calculation of several Boolean functions at once. Another useful property
of these arithmetic representations is used for linearization techniques.

\section{\label{sec:Elective-projector-logic}Elective projector logic}

The following section presents the real new part of this work. It
will be shown that the results given above can be applied within the
framework of the following formalism. It must be emphasized that at
the time of George Boole methods in matrix linear algebra were in
their nascent form. Most methods have been introduced around 1850,
major contributions are due to Arthur Cayley and James Joseph Sylvester,
the latter having introduced the term \textit{matrix}. The modern
definition of a vector space was subsequently introduced by Giuseppe
Peano in 1888.

\subsection{Parallels to Boole's expansion with idempotent operators in linear
algebra}

One question arises: why one would want to find parallels to Boole's
expansion theorem for idempotent functions of idempotent symbols in
linear algebra? One of the principal motivations of this work is seeking
the links with operational algebra as is used in quantum mechanics
in Hilbert space with applications in the emerging field of \textit{quantum
information} and \textit{quantum computation} \cite{key-9}.

Concerning the possible applications of the idempotent linear operator
algebra version of Boole's operator algebra to quantum mechanics,
some important things can be recalled. Quantum mechanics was a hot
topic at Harvard starting in the late 1920s. Marshall H. Stone, a
student of Garret D. Birkhoff, wrote a book in the early 1930's on
linear operators on infinite dimensional spaces \cite{key-61} he
then subsequently, starting in 1934, undertook a great research effort
in logic culminating in two papers on Boolean algebras, Boolean rings,
and Boolean spaces \cite{key-71,key-72}.

Marshall H. Stone showed that any Boolean algebra is isomorphic to
a field of sets, and he motivated his algebraic approach to logic
by the fact that it allows to connect many different areas of mathematics.
As underlined by Stanley Burris \cite{key-21} it is interesting to
note that his motivation for studying Boolean algebra came from the
mathematics of areas like quantum mechanics: (quote from his 1936
paper \cite{key-71}) ``The writer's interest in the subject, for
example, arose in connection with the spectral theory of symmetric
transformations in Hilbert space and certain related properties of
abstract integrals.\textquotedblright{} This could have meant that
he was looking at Boolean algebras of idempotent linear transformations,
and realized that there were a lot of examples of Boolean algebras
that had not been considered before. He goes on to prove that Huntington's
axiomatization of Boolean algebras \cite{key-11} is equivalent with
the axiomatization of commutative rings with unit element, in which
every element is idempotent called Boolean rings (\cite{key-71} p.
38).

According to Dirk Schlimm in \cite{key-81} Marshall H. Stone was
able to connect the theory of Boolean rings also to topology by proving
that ``the theory of Boolean rings is mathematically equivalent to
the theory of locally-bicompact totally-disconnected topological spaces\textquotedblright .
This identification, also referred to as the \textit{fundamental representation
theorem} allows for the transfer of topological methods to the study
of Boolean algebras, and vice-versa, is known as the \textit{Stone
duality}.

There has also been work on developing a specific logic for quantum
mechanics by Garrett Birkhoff and John von Neumann in their 1936 seminal
paper on the subject \cite{key-91}, they proposed the replacement
of Boolean algebras with the lattice of closed subspaces of a (finite)
Hilbert space. Quantum logic has become an independent discipline
with many promoters and different versions, even though it has not
still reached the status of an ``operational tool\textquotedblright{}
in the emerging quantum information and quantum computing fields.
Already in 1932 John von Neumann made parallels between projections
in Hilbert space and logical propositions (p. 249: ``Projectors as
Propositions'' in \cite{key-92}). As is clearly stated by François
David in \cite{key-93} John von Neumann noticed that the \textit{observables}
(name given to Hermitian operators in quantum mechanics) given by
projection operators $\mathbf{P}$, such that $\mathbf{P}^{2}=\mathbf{P}=\mathbf{P}^{\dagger}$,
correspond to propositions with a \textit{Yes} or \textit{No} (\textit{i.e.}
\textit{True} or \textit{False}) outcome in a logical system.

An orthogonal projection operator $\mathbf{P}$ onto a linear subspace
$P$, in Hilbert space, is indeed an observable that can take only
the eigenvalues $1$ (if the corresponding quantum state belongs the
subspace $P$) or $0$ (if the corresponding quantum state belongs
to the orthogonal subspace to $P$). Thus the two values $1$ and
$0$ are the only possible eigenvalues of the projection operator
$\mathbf{P}$, and this statement, that a measurement can only give
one of the eigenvalues, is part of the fundamental measurement postulate
in Quantum Mechanics \cite{key-92,key-93,key-94}. Thus measuring
the observable $\mathbf{P}$ is equivalent to perform a test on the
system, or to check the validity of a logical proposition on the system,
which can only be true or false, and not some combination of these
values. This states in other terms the Aristotelian \textit{law of
the excluded middle} for a proposition.

In his 1932 book \cite{key-92} John von Neumann cites the book of
Marshall H. Stone (p.70: ``Projections\textquotedblright{} in \cite{key-61})
about the operations conserving the properties of projection operators
and gives the following rules: 
\begin{itemize}
\item $\mathbf{\qquad P}_{1}\cdot\mathbf{P}_{2}\quad$ is a projection operator
if and only if $\mathbf{\quad P}_{1}\cdot\mathbf{P}_{2}\equiv\mathbf{P}_{2}\cdot\mathbf{P}_{1}\quad$
(\textit{i.e.} they commute) 
\item $\mathbf{\qquad P}_{1}+\mathbf{P}_{2}\quad$ is a projection operator
if and only if $\mathbf{\quad P}_{1}\cdot\mathbf{P}_{2}\equiv0\quad$
or $\quad\mathbf{P}_{2}\cdot\mathbf{P}_{1}\equiv0$ 
\item $\mathbf{\qquad P}_{1}-\mathbf{P}_{2}\quad$ is a projection operator
if and only if $\mathbf{\quad P}_{1}\cdot\mathbf{P}_{2}\equiv\mathbf{P}_{2}\quad$
or $\mathbf{\quad P}_{2}\cdot\mathbf{P}_{1}\equiv\mathbf{P}_{2}$ 
\end{itemize}
this shows that the property of projection operators, i.e. idempotence,
is conserved under the operations of (matrix) product $\mathbf{P}_{1}\cdot\mathbf{P}_{2}$,
sum $\mathbf{P}_{1}+\mathbf{P}_{2}$ and difference $\mathbf{P}_{1}-\mathbf{P}_{2}$
only for commuting projection operators, this condition is usually
expressed in quantum mechanics by the commutation relation $\mathbf{P}_{1}\cdot\mathbf{P}_{2}-\mathbf{P}_{2}\cdot\mathbf{P}_{1}=\left[\mathbf{P}_{1},\mathbf{P}_{2}\right]=0$
. The sum is only defined for disjoint subspaces, $P_{1}\cap P_{2}\equiv0$,
and the difference with the inclusion of subspaces $P_{2}\subseteq P_{1}$.
These properties will be at the basis of the development given hereafter
for \textit{Eigenlogic}, establishing the connection between eigenvalues
and logic because of the fact that idempotent diagonal matrices have
only $1$'s and $0$'s on the diagonal, and hence these are the only
possible outcomes (eigenvalues).

Also it is interesting to note that the very definition of a pure
quantum state when expressed by a density matrix, also introduced
by John von Neumann, is a \textit{ray} (a rank-1 idempotent projection
operator spanning a one-dimensional subspace). All these concepts
lay at the foundations of quantum theory.

The work presented here can be understood in this framework, even
though one does not need here (at least at this stage) the non-commutative
algebra which is at the basis of the peculiar aspects of quantum theory,
having as consequence, for example, the non-distributivity of quantum
logic. The approach here can be viewed as \textit{classical} in the
sense that the discussion is restricted to families of commuting observables
which are here projection operators. But because this approach uses
observables it can also be considered as being part of the global
``quantum machinery''. Most problems in traditional quantum physics
deal with finding eigenfunctions and eigenvalues of some physical
observable, the most investigated being the Hamiltonian observables
whose eigenvalues represent the energies of a physical system and
whose eigenstates are the stationary states representing the stable
equilibrium solutions, in the form of wavefunctions, of the Schrödinger
equation. The non-traditional aspects of quantum mechanics, principally
superposition, entanglement and non-commutativity, are largely employed
in the field of quantum information and are considered as a resource
for quantum computing \cite{key-94}. Nothing in the formulation presented
here forbids to explore outside of the family of commuting logical
projection operators, or to consider vectors that are not eigenvectors
of the same logical family. This is the object of ongoing research
(see \cite{key-9}).

\subsection{Link of George Boole's formulation and linear algebra}

If one goes back to the motivation of George Boole's elective symbols,
one sees that he applies them as selecting operators on classes of
objects. As outlined in \cite{key-3} expressions which do not represent
classes are called by George Boole ``uninterpretable\textquotedblright ,
and are formally recognizable as those which do not satisfy the idempotence
law $x^{2}=x$. Characteristic of the method is that while expressions
may be uninterpretable, equations always are when suitably interpreted
by rules.

But in his first book \cite{key-1} he was limited by the interpretation
of the number $1$ which he considered as the unique class $U$ representing
the whole universe. Because of this, without going into all the details,
see for example \cite{key-4,key-5}, he changed the method in his
second book in 1854 \cite{key-2} and applied the formalism to subclasses
of the universal class $U$.

Modern terminology will be used to describe what George Boole was
doing: the word \textit{class} should be used as a synonym for the
modern word \textit{set}. In \cite{key-1} he starts with the universe
class $U$ and looks successively in \cite{key-2} at the collection
$P(U)$ of subclasses. The definition of the selection (\textit{i.e.}
\textit{elective}) operator $S_{A}$ defined for $P(U)\rightarrow P(U)$
for $A\in P(U)$ acting for $X\in P(U)$ is given, by the intersection:

\begin{equation}
S_{A}(X)\coloneqq A\cap X\label{eq:Selective-operator}
\end{equation}

Using composition of operators for multiplication, his operators were
associative, commutative and idempotent. Letting $0$ be the empty
class, $1$ the universe $U$, one has $S_{0}(X)=0$, $S_{1}(X)=X$.
Addition was partially defined, namely $S_{A}+S_{B}$, was defined
for $A\cap B=0$. Likewise subtraction was also partially defined.

When considering all the laws that George Boole actually uses $(+,-,\times,0,1)$
viewed as a set of axioms for a mathematical theory, Theodore Hailperin
finds \cite{key-3} that the correct interpretations or models are
obtained if one considers, not classes, but multisets as the entities
over which the variables range. The operators defined here-above carry
over to \textit{signed multisets}, which are conveniently expressed
as a map $f\,:\,U\rightarrow\mathbb{Z}$. Then George Boole's classes
correspond to characteristic functions by the means of the map $\alpha\,:\,\Lambda\rightarrow\hat{\Lambda}$,
where $\hat{\Lambda}(u)$ is $1$ if $u\in\Lambda$ and $0$ otherwise.
The collection of maps from $U$ to $\mathbb{Z}$ is usually written
as $\mathbb{Z}{}^{U}$, a ring of functions with scalar multiplication
(by elements of $\mathbb{Z}$), where the operations are given pointwise,
that is, for $u\in U$. Boole's election operators $S_{A}$ on $P(U)$
can thus be translated to corresponding operators which are the set
of idempotent elements of the ring $\mathbb{Z}{}^{U}$.

If one wants to use linear operations on a vector space, one needs
to extend the ring $\mathbb{Z}{}^{U}$ to a field $F$, since vector
spaces are defined over fields, thus the set of idempotents $\{0,1\}^{U}$,
the ring of signed multisets $\mathbb{Z}{}^{U}$ and the algebra of
functions $F^{U}$ over $F$ verify:

\begin{equation}
\{0,1\}^{U}\subseteq\mathbb{Z}{}^{U}\subseteq F^{U}\label{eq:Ring Field}
\end{equation}

The isomorphism between the ring $\mathbb{Z}{}^{U}$ restricted to
its idempotent elements $\{0,1\}^{U}$ and Boole's algebra of classes
on $P(U)$ is due to Theodore Hailperin in \cite{key-3}. His breakthrough
was to point out this equivalence: the set of elements $x$ of an
algebra of signed multisets which satisfy $x^{2}=x$ constitute a
Boolean algebra. But most importantly all the axioms that were needed
by Boole's (partial) algebra of logic hold in the complete algebra
$\mathbb{Z}{}^{U}$. This means that Boole's equational reasoning
was correct in $\mathbb{Z}{}^{U}$, and thus in his partial algebra
$P(U)$. So finally, as is pointed out by Stanley Burris \cite{key-21},
much of Boole's work in logic had a solid foundation.

There is also an isomorphism between the ring of linear operators
on $F^{U}$, restricted to those linear operators defined by left
multiplication (i.e. ordered matrix product) by an idempotent element
of $F^{U}$ and Boole's algebra of selection operators $S_{A}$ on
$P(U)$. A linear operator on $F^{U}$ that is defined by left multiplication
by an idempotent is the same as the one given by left multiplication
by a diagonal matrix with the idempotent characteristic function $\hat{\Lambda}$
along the diagonal.

From Theodore Hailperin's book \cite{key-4} it is clear that given
any commutative ring $R$ with unity and without nilpotent elements
one has parallels to all of George Boole's theorems, not just the
development theorem, holding in the ring. One can think of such a
ring as a ring of operators acting by left multiplication on $R$.
Indeed $R$ can be viewed as a unitary left $R$-module. Thus one
also has parallels to Boole's results in \cite{key-1}.

If one takes the ring $R$ to be the ring $\mathbb{Z}{}^{N}$ of $N$-tuples
of integers, then the idempotent elements are the $N$-tuples with
$\left\{ 0,1\right\} $ entries. By identifying the $N$-tuple operators
with $N\times N$ diagonal matrices (vector space of dimension $d=N$),
and the elements of the ring with column vectors, one gets the linear
algebra situation treated hereafter. It must be outlined that because
of binary cardinality we have here $d=N=2^{n}$.

\subsection{The seed projector and one argument operators}

As stated above the elective symbols represent operators acting on
a given class of objects (a subclass $P(U)$ of the universe class
$U$). In this way the elective operator represented by the number
$1$ will simply become the identity operator for the considered subclass.
Using the framework of linear algebra, operators are defined on a
vector space whose dimension depends on the number of arguments (the
arity) in the propositional system.

So what operators can represent the selection of elements out of a
class? The straightforward answer in linear algebra are the projection
operators which have the property of idempotence.

Considering the case of objects belonging to one single class, the
corresponding projection operator $\boldsymbol{\Pi}$ of this class
will act on vectors. Now what are the expected outcomes when applying
this projector? If a vector $\overrightarrow{(a)}$ corresponds exactly
to elements of the class, the following matrix equations will be verified:

\begin{equation}
\boldsymbol{\Pi}_{(1)}\cdot\overrightarrow{(a)}=1\cdot\overrightarrow{(a)}\qquad\qquad\qquad\boldsymbol{\Pi}_{(0)}\cdot\overrightarrow{(a)}=0\cdot\overrightarrow{(a)}\label{eq:EigenEq2D}
\end{equation}

The values $0$ and $1$ are the two eigenvalues of the two projectors
associated with the eigenvector $\overrightarrow{(a)}$. As before,
if interpretable results are to be considered in logic, the only possible
numbers for these eigenvalues are $0$ and $1$. $1$ will be obtained
for objects belonging to the considered class and $0$ for objects
not belonging to it. In the second case one can also define the complement
vector $\overrightarrow{(\overline{a})}$.

The \textit{True} eigenvalue $1$ will correspond to the eigenvector
$\overrightarrow{(a)}$, named $\overrightarrow{(1)}$, and the\textit{
False} eigenvalue $0$ will correspond to the complementary eigenvector
$\overrightarrow{(\overline{a})}$ named $\overrightarrow{(0)}$.

When these properties are expressed in matrix form the projection
operators $\boldsymbol{\Pi}_{(1)}$ and $\boldsymbol{\Pi}_{(0)}$
are $2\times2$ square matrices and the vectors $\overrightarrow{(a)}$
and $\overrightarrow{(\overline{a})}$ are $2$ dimensional orthonormal
column vectors:

\begin{equation}
\boldsymbol{\Pi}_{(1)}=\boldsymbol{\Pi}=\left(\begin{array}{cc}
0 & 0\\
0 & 1
\end{array}\right)\qquad\qquad\boldsymbol{\Pi}_{(0)}=\boldsymbol{\mathbf{I}}_{2}-\boldsymbol{\Pi}=\left(\begin{array}{cc}
1 & 0\\
0 & 0
\end{array}\right)\label{eq:projPi}
\end{equation}

\begin{equation}
\overrightarrow{(a)}=\overrightarrow{(1)}=\left(\begin{array}{c}
0\\
1
\end{array}\right)\qquad\qquad\overrightarrow{(\overline{a})}=\overrightarrow{(0)}=\left(\begin{array}{c}
1\\
0
\end{array}\right)\label{eq:eigenvector2d}
\end{equation}

The two projectors given in equation (\ref{eq:projPi}) are complementary
and idempotent, this last condition is written:

\begin{equation}
\boldsymbol{\Pi}\cdot\boldsymbol{\Pi}=\boldsymbol{\Pi}^{2}=\boldsymbol{\Pi}\label{eq:IdempotProj}
\end{equation}

One can then construct the $4$ logical operators corresponding to
the $4$ elective functions given in table \ref{Table 1} corresponding
to the single argument case $n=1$. Capital bold letters are used
here to represent operators.

\begin{eqnarray}
\boldsymbol{A} & = & \boldsymbol{\Pi}=\left(\begin{array}{cc}
0 & 0\\
0 & 1
\end{array}\right)\qquad\qquad\boldsymbol{\bar{A}}=\boldsymbol{\mathbf{I}}_{2}-\boldsymbol{\Pi}=\left(\begin{array}{cc}
1 & 0\\
0 & 0
\end{array}\right)\nonumber \\
 &  & \boldsymbol{True}=\boldsymbol{\mathbf{I}}_{2}=\left(\begin{array}{cc}
1 & 0\\
0 & 1
\end{array}\right)\qquad\qquad\boldsymbol{False}=\boldsymbol{0}_{2}=\left(\begin{array}{cc}
0 & 0\\
0 & 0
\end{array}\right)\label{eq:LogOperator1}
\end{eqnarray}

$\boldsymbol{A}$ is the \textit{logical projector} and $\boldsymbol{\bar{A}}$
its complement. The $\boldsymbol{True}$ (tautology) operator corresponds
here to the identity operator in $2$ dimensions $\boldsymbol{I}_{2}$
. $\boldsymbol{False}$ (contradiction) corresponds here to the nil
operator $\boldsymbol{0}_{2}$.

Remark that $\boldsymbol{I}_{2}$ and $\boldsymbol{0}_{2}$ are also
projection operators (idempotent). So in general for one argument
the matrix form of the projection operator corresponding to the logical
function $f_{i}^{[1]}(x)$ gi$R$ ven on Table \ref{Table 1} is:

\begin{equation}
\boldsymbol{F}_{i}^{[1]}=f_{i}^{[1]}(0)\:\boldsymbol{\Pi}_{(0)}+f_{i}^{[1]}(1)\:\boldsymbol{\Pi}_{(1)}=\left(\begin{array}{cc}
f_{i}^{[1]}(0) & 0\\
0 & f_{i}^{[1]}(1)
\end{array}\right)\label{eq:Matrix-one-arg}
\end{equation}

This equation represents the \textit{spectral decomposition} of the
operator and because the eigenvalues are real the logical operator
is Hermitian and can thus be considered as an observable. In this
way, in Eigenlogic, the truth values of the logical proposition are
the eigenvalues of the logical observable. In the very simple case
where 0 and 1 are both not degenerate eigenvalues, the projection
operators relative to the eigenvector basis take the form of the logical
projector $\boldsymbol{A}$ and its complement $\boldsymbol{\bar{A}}$.
As is done in quantum mechanics one can find the set of projection
operators that completely represent the system, in particular by lifting
the eventual degeneracy of the eigenvalues. Here eigenvalues are always
equal to 0 or 1 and the question about the multiplicity of eigenvalues
is natural. This last point is important in the model, because not
only mutually exclusive projection operators are representative of
a logical system, the complete family of commuting projection operators
(the \textit{logical family}) must be used in order to completely
define the logical system. When these properties are expressed in
matrix terms this means that the matrix product of the logical observables
are not necessarily $0$.

\subsection{Extending to more arguments}

As seen above when representing logic with $n$ arguments ($n$-arity)
using idempotent projection operators various possibilities are intrinsically
present in a unique structure with $\,2^{2^{n}}\,$ different projection
operators. Once the eigenbasis is chosen the remaining structure is
intrinsic thus basis independent.

The extension to more arguments can be obtained by increasing the
dimension, this is done by using the Kronecker product $\otimes$.
It is a standard procedure in linear algebra justified because it
can be shown (Wedderburn little theorem \cite{key-93}) that any finite
division ring (a divison ring is the analogue of a field without necessitating
commutativity) is a direct product of Galois fields $GF_{p}=\mathbb{Z}/\mathbb{Z}_{p}$
($p$ prime), in the binary case considered here $p=2$. The direct
product becomes explicitly the tensor or Kronecker product of linear
operators.

In this work the application of this method was originally inspired
from the composition rule of quantum states, which has the status
nowadays of postulate in quantum mechanics \cite{key-94} where the
quantum state vector corresponding to the composition of two quantum
systems represented by two subspaces in Hilbert space, is the Kronecker
product of the respective quantum state vectors. The operators acting
in the combined space are combinations of the quantum operators in
the respective sub-spaces. The interesting fact is that for the combined
case new structures appear, named \textit{non-local}, that cannot
be put as simple Kronecker products but are linear combinations of
these. It will be shown that several projection operators presented
hereafter corresponding to logical observables are not simply Kronecker
products of elementary projection operators.

In the following, as before for the elective logical functions, superscripts
are used in order to indicate how many arguments are used (arity)
in the propositional system.

One can verify that in equation (\ref{eq:LogOperator1}) all the four
logical operators are effectively idempotent and commuting. The correspondence
of the elective symbol $x$ with the elementary \textit{seed} projector
$\boldsymbol{\Pi}$ will be used in the following to build higher
arity logical operators.

For $2$ arguments (arity $n=2$) one needs $4$ commuting orthogonal
rank-$1$ projector operators in order to express the development
in the same way as in equation (\ref{eq:ElectDev2}).

Some properties of the Kronecker product on idempotent projection
operators have to be outlined. 
\begin{lyxlist}{00.00.0000}
\item [{(i)}] The Kronecker product of two projection operators is also
a projection operator. 
\item [{(ii)}] If projection operators are rank-$1$ (a single eigenvalue
is $1$ all the others are $0$) then their Kronecker product is also
a rank-$1$ projection operator. 
\end{lyxlist}
Using these two properties, the $4$ commuting orthogonal rank-$1$
projectors spanning the $4$ dimensional vector space can be calculated
in a straightforward way:\\
 
\[
\boldsymbol{\Pi}_{(0,0)}^{[2]}=(\boldsymbol{\mathbf{I}}_{2}-\boldsymbol{\Pi})\otimes(\boldsymbol{I}_{2}-\boldsymbol{\Pi})=\left(\begin{array}{cccc}
1 & 0 & 0 & 0\\
0 & 0 & 0 & 0\\
0 & 0 & 0 & 0\\
0 & 0 & 0 & 0
\end{array}\right)\qquad\boldsymbol{\Pi}_{(0,1)}^{[2]}=(\boldsymbol{\mathbf{I}}_{2}-\boldsymbol{\Pi})\otimes\boldsymbol{\Pi}=\left(\begin{array}{cccc}
0 & 0 & 0 & 0\\
0 & 1 & 0 & 0\\
0 & 0 & 0 & 0\\
0 & 0 & 0 & 0
\end{array}\right)
\]

\begin{equation}
\boldsymbol{\Pi}_{(1,0)}^{[2]}=\boldsymbol{\Pi}\otimes(\boldsymbol{\mathbf{I}}_{2}-\boldsymbol{\Pi})=\left(\begin{array}{cccc}
0 & 0 & 0 & 0\\
0 & 0 & 0 & 0\\
0 & 0 & 1 & 0\\
0 & 0 & 0 & 0
\end{array}\right)\qquad\boldsymbol{\Pi}_{(1,1)}^{[2]}=\boldsymbol{\Pi}\otimes\boldsymbol{\Pi}=\left(\begin{array}{cccc}
0 & 0 & 0 & 0\\
0 & 0 & 0 & 0\\
0 & 0 & 0 & 0\\
0 & 0 & 0 & 1
\end{array}\right)\label{eq:OrthProj2}
\end{equation}

By the same procedure as in equation (\ref{eq:ElectDev2}), one can
write the operators for $n=2$ arguments for a two-argument function
(see Table \ref{Table 2}) using the projectors given in equation
(\ref{eq:OrthProj2}): $R$
\begin{equation}
\boldsymbol{F}_{i}^{[2]}=f_{i}^{[2]}(0,0)\:\boldsymbol{\Pi}_{(0,0)}^{[2]}+f_{i}^{[2]}(0,1)\:\boldsymbol{\Pi}_{(0,1)}^{[2]}+f_{i}^{[2]}(1,0)\:\boldsymbol{\Pi}_{(1,0)}^{[2]}+f_{i}^{[2]}(1,1)\:\boldsymbol{\Pi}_{(1,1)}^{[2]}\label{eq:OperatDecomp2}
\end{equation}

\begin{equation}
\boldsymbol{F}_{i}^{[2]}=\left(\begin{array}{cccc}
f_{i}^{[2]}(0,0) & 0 & 0 & 0\\
0 & f_{i}^{[2]}(0,1) & 0 & 0\\
0 & 0 & f_{i}^{[2]}(1,0) & 0\\
0 & 0 & 0 & f_{i}^{[2]}(1,1)
\end{array}\right)\label{eq:MatrixDecomp2-1}
\end{equation}

The coefficients (co-factors) are the logical function's truth values
given on table \ref{Table 2}.

This method can be extended to whatever number of arguments $n$ using
the same seed projector $\boldsymbol{\Pi}$ and its complement $(\boldsymbol{\mathbf{I}}_{2}-\boldsymbol{\Pi})$.

\subsection{Logical observables for two arguments}

For arity $n=2$ the polynomial expressions have already been calculated
in table \ref{Table 2}, so one can write down directly the corresponding
operators. One has to express the logical projectors corresponding
to the two arguments $x=a$ and $y=b$ and this is given using equation
(\ref{eq:OperatDecomp2}) by considering the truth values of the functions
$f_{12}^{[2]}$ and $f_{10}^{[2]}$, these operators are:

\begin{equation}
\boldsymbol{A}^{[2]}=\boldsymbol{F}_{12}^{[2]}=1\cdot\boldsymbol{\Pi}_{(1,0)}^{[2]}+1\cdot\boldsymbol{\Pi}_{(1,1)}^{[2]}=\boldsymbol{\Pi}\otimes(\mathbf{\mathbf{I}}_{2}-\boldsymbol{\Pi})+\boldsymbol{\Pi}\otimes\boldsymbol{\Pi}=\boldsymbol{\Pi}\otimes\mathbf{\mathbf{I}}_{2}=\left(\begin{array}{cccc}
0 & 0 & 0 & 0\\
0 & 0 & 0 & 0\\
0 & 0 & 1 & 0\\
0 & 0 & 0 & 1
\end{array}\right)\label{eq: LogProj2A}
\end{equation}

\begin{equation}
\boldsymbol{B}^{[2]}=\boldsymbol{F}_{10}^{[2]}=1\cdot\boldsymbol{\Pi}_{(0,1)}^{[2]}+1\cdot\boldsymbol{\Pi}_{(1,1)}^{[2]}=(\mathbf{\mathbf{I}}_{2}-\boldsymbol{\Pi})\otimes\boldsymbol{\Pi}+\boldsymbol{\Pi}\otimes\boldsymbol{\Pi}=\mathbf{\mathbf{I}}_{2}\otimes\boldsymbol{\Pi}=\left(\begin{array}{cccc}
0 & 0 & 0 & 0\\
0 & 1 & 0 & 0\\
0 & 0 & 0 & 0\\
0 & 0 & 0 & 1
\end{array}\right)\label{eq:LogProj2B}
\end{equation}

Here are some examples: the conjunction operator for $n=2$ will simply
be the product of the two logical projectors:

\begin{equation}
\boldsymbol{F}_{AND}^{[2]}=\boldsymbol{A}^{[2]}\cdot\boldsymbol{B}^{[2]}=(\boldsymbol{\Pi}\otimes\mathbf{\mathbf{I}}_{2})\cdot(\mathbf{\mathbf{I}}_{2}\otimes\boldsymbol{\Pi})=\boldsymbol{\Pi}\otimes\boldsymbol{\Pi}=\left(\begin{array}{cccc}
0 & 0 & 0 & 0\\
0 & 0 & 0 & 0\\
0 & 0 & 0 & 0\\
0 & 0 & 0 & 1
\end{array}\right)\label{eq:OperAND2}
\end{equation}

where the following property of the Kronecker product has been used:
if $\boldsymbol{P}$, $\boldsymbol{Q}$, $\boldsymbol{R}$ and $\boldsymbol{S}$
are operators then:

\begin{equation}
(\boldsymbol{P}\otimes\boldsymbol{Q})\cdot(\boldsymbol{R}\otimes\boldsymbol{S})=(\boldsymbol{P}\cdot\boldsymbol{R})\otimes(\boldsymbol{Q}\cdot\boldsymbol{S})\label{eq:KroneckerTrick}
\end{equation}

The disjunction operator can be directly written, using equation (\ref{eq:electiveOR}):
\begin{equation}
\boldsymbol{F}_{OR}^{[2]}=\boldsymbol{A}^{[2]}+\boldsymbol{B}^{[2]}-\boldsymbol{A}^{[2]}\cdot\boldsymbol{B}^{[2]}=\left(\begin{array}{cccc}
0 & 0 & 0 & 0\\
0 & 1 & 0 & 0\\
0 & 0 & 1 & 0\\
0 & 0 & 0 & 1
\end{array}\right)
\end{equation}
\\

The exclusive disjunction can also be directly written, using equation
(\ref{eq:electiveXOR}):

\begin{equation}
\boldsymbol{F}_{XOR}^{[2]}=\boldsymbol{A}^{[2]}+\boldsymbol{B}^{[2]}-2\boldsymbol{A}^{[2]}\cdot\boldsymbol{B}^{[2]}=\left(\begin{array}{cccc}
0 & 0 & 0 & 0\\
0 & 1 & 0 & 0\\
0 & 0 & 1 & 0\\
0 & 0 & 0 & 0
\end{array}\right)
\end{equation}

Negation is obtained by subtracting from the identity operator (complementation)
giving in general for $n$ arguments:

\begin{eqnarray}
\boldsymbol{\bar{A}}^{[n]} & = & \mathbf{\mathbf{I}}_{2^{n}}-\boldsymbol{A}^{[n]}\label{eq:OperatorNeg}
\end{eqnarray}

this equation can be used to obtain the $NAND$ operator:

\begin{equation}
\boldsymbol{F}_{NAND}^{[2]}=\mathbf{\mathbf{I}}_{4}-\boldsymbol{F}_{AND}^{[2]}=\mathbf{\mathbf{I}}_{4}-\boldsymbol{A}^{[2]}\cdot\boldsymbol{B}^{[2]}
\end{equation}

Using De Morgan's law:

\begin{equation}
\boldsymbol{F}_{NOR}^{[2]}=(\mathbf{\mathbf{I}}_{4}-\boldsymbol{A}^{[2]})\cdot(\mathbf{\mathbf{I}}_{4}-\boldsymbol{B}^{[2]})=\mathbf{\mathbf{I}}_{4}-\boldsymbol{A}^{[2]}-\boldsymbol{B}^{[2]}+\boldsymbol{A}^{[2]}\cdot\boldsymbol{B}^{[2]}=\left(\begin{array}{cccc}
1 & 0 & 0 & 0\\
0 & 0 & 0 & 0\\
0 & 0 & 0 & 0\\
0 & 0 & 0 & 0
\end{array}\right)=(\mathbf{\mathbf{I}}_{2}-\boldsymbol{\Pi})\otimes(\mathbf{\mathbf{I}}_{2}-\boldsymbol{\Pi})
\end{equation}

Material implication is also straightforwardly obtained using the
expression given in Table \ref{Table 2}:

\begin{equation}
\boldsymbol{F}_{\Rightarrow}^{[2]}=\mathbf{\mathbf{I}}_{4}-\boldsymbol{A}^{[2]}+\boldsymbol{A}^{[2]}\cdot\boldsymbol{B}^{[2]}=\left(\begin{array}{cccc}
1 & 0 & 0 & 0\\
0 & 1 & 0 & 0\\
0 & 0 & 0 & 0\\
0 & 0 & 0 & 1
\end{array}\right)=\mathbf{\mathbf{I}}_{4}-(\boldsymbol{\Pi})\otimes(\mathbf{\mathbf{I}}_{2}-\boldsymbol{\Pi})
\end{equation}

On table \ref{Tab3} are given the logical operator forms for the
$16$ two-argument logical connectives.

\begin{table}
\centering{}%
\begin{tabular}{|c|c|c|c|}
\hline 
connective for  & operator diagonal  & logical  & logical\tabularnewline
Boolean  & form  & observable $\boldsymbol{F}_{i}^{[2]}$  & observable $\boldsymbol{F}_{i}^{[2]}$\tabularnewline
$A,B$  & $diag(\mathbf{truth\:vector})$  & $\boldsymbol{A\:,\:}\boldsymbol{B}$ argument form  & $\boldsymbol{\Pi}$ seed operator form \tabularnewline
\hline 
\hline 
False F  & $diag(0,0,0,0)$  & $\boldsymbol{\mathit{\mathbf{0}}}$  & $\boldsymbol{\mathit{\mathbf{0}}}$\tabularnewline
\hline 
NOR ; $\overline{A\vee B}$  & $diag(1,0,0,0)$  & $\mathbf{\mathbf{I}}-\boldsymbol{A}-\boldsymbol{B}+\boldsymbol{A}\cdot\boldsymbol{B}$  & $\left(\mathbf{\mathbf{I}}-\boldsymbol{\Pi}\right)\otimes\left(\mathbf{\mathbf{I}}-\boldsymbol{\Pi}\right)$ \tabularnewline
\hline 
$A\,\nLeftarrow\,1B$  & $diag(0,1,0,0)$  & $\boldsymbol{B}-\boldsymbol{A}\cdot\boldsymbol{B}$  & $\boldsymbol{\Pi}\otimes\left(\mathbf{\mathbf{I}}-\boldsymbol{\Pi}\right)$ \tabularnewline
\hline 
$\overline{A}$  & $diag(1,1,0,0)$  & $\mathbf{\mathbf{\mathbf{I}}}-\boldsymbol{A}$  & $\mathbf{\mathbf{I}}-\left(\boldsymbol{\Pi}\otimes\mathbf{\mathbf{I}}\right)$\tabularnewline
\hline 
$A\nRightarrow B$  & $diag(0,0,1,0)$  & $\boldsymbol{A}-\boldsymbol{A}\cdot\boldsymbol{B}$  & $\left(\mathbf{\mathbf{I}}-\boldsymbol{\Pi}\right)\otimes\boldsymbol{\Pi}$ \tabularnewline
\hline 
$\overline{B}$  & $diag(1,0,1,0)$  & $\mathbf{\mathbf{\mathbf{I}}}-\boldsymbol{B}$  & $\mathbf{\mathbf{I}}-\left(\mathbf{\mathbf{I}}\otimes\boldsymbol{\Pi}\right)$ \tabularnewline
\hline 
$A\oplus B$  & $diag(0,1,1,0)$  & $\boldsymbol{A}+\boldsymbol{B}-2\boldsymbol{A}\cdot\boldsymbol{B}$  & $\boldsymbol{\Pi}\otimes\left(\mathbf{\mathbf{I}}-\boldsymbol{\Pi}\right)+\left(\mathbf{\mathbf{I}}-\boldsymbol{\Pi}\right)\otimes\boldsymbol{\Pi}$ \tabularnewline
\hline 
NAND ; $\overline{A\wedge B}$  & $diag(1,1,1,0)$  & $\mathbf{\mathbf{I}}-\boldsymbol{A}\cdot\boldsymbol{B}$  & $\mathbf{\mathbf{I}}-\left(\boldsymbol{\Pi}\otimes\boldsymbol{\Pi}\right)$\tabularnewline
\hline 
AND ; $A\wedge B$  & $diag(0,0,0,1)$  & $\boldsymbol{A}\cdot\boldsymbol{B}$  & $\boldsymbol{\Pi}\otimes\boldsymbol{\Pi}$ \tabularnewline
\hline 
$A\equiv B$  & $diag(1,0,0,1)$  & $\mathbf{\mathbf{I}}-\boldsymbol{A}-\boldsymbol{B}+2\boldsymbol{A}\cdot\boldsymbol{B}$  & $\boldsymbol{\Pi}\otimes\boldsymbol{\Pi}+\left(\mathbf{\mathbf{I}}-\boldsymbol{\Pi}\right)\otimes\left(\mathbf{\mathbf{I}}-\boldsymbol{\Pi}\right)$ \tabularnewline
\hline 
$B$  & $diag(0,1,0,1)$  & $\boldsymbol{B}$  & $\mathbf{\mathbf{I}}\otimes\boldsymbol{\Pi}$\tabularnewline
\hline 
$A\Rightarrow B$  & $diag(1,1,0,1)$  & $\mathbf{\mathbf{\mathbf{I}}}-\boldsymbol{A}+\boldsymbol{A}\cdot\boldsymbol{B}$  & $\mathbf{\mathbf{I}}-\left[\left(\mathbf{\mathbf{I}}-\boldsymbol{\Pi}\right)\otimes\boldsymbol{\Pi}\right]$ \tabularnewline
\hline 
$A$  & $diag(0,0,1,1)$  & $\boldsymbol{A}$  & $\boldsymbol{\Pi}\otimes\mathbf{\mathbf{I}}$ \tabularnewline
\hline 
$A\Leftarrow B$  & $diag(1,0,1,1)$  & $\mathbf{\mathbf{I}}-\boldsymbol{B}+\boldsymbol{A}\cdot\boldsymbol{B}$  & $\mathbf{\mathbf{I}}-\left[\boldsymbol{\Pi}\otimes\left(\mathbf{\mathbf{I}}-\boldsymbol{\Pi}\right)\right]$\tabularnewline
\hline 
OR ; $A\vee B$  & $diag(0,1,1,1)$  & $\boldsymbol{A}+\boldsymbol{B}-\boldsymbol{A}\cdot\boldsymbol{B}$  & $\mathbf{\mathbf{I}}-\left[\left(\mathbf{\mathbf{I}}-\boldsymbol{\Pi}\right)\otimes\left(\mathbf{\mathbf{I}}-\boldsymbol{\Pi}\right)\right]$ \tabularnewline
\hline 
True T  & $diag(1,1,1,1)$  & $\mathbf{I}$  & $\mathbf{I}$ \tabularnewline
\hline 
\end{tabular}\caption{The sixteen two-argument connectives and the respective Eigenlogic
logic observables }
\label{Tab3} 
\end{table}

\subsection{Logical observables for three arguments}

For arity $n=3$ one can generate $8$ orthogonal $8$-dimensional
rank-$1$ projectors, for example two of these are given, by:\\

\begin{equation}
\boldsymbol{\Pi}_{(1,1,1)}^{[3]}=\boldsymbol{\Pi}\otimes\boldsymbol{\Pi}\otimes\boldsymbol{\Pi}\qquad\qquad\boldsymbol{\Pi}_{(0,1,0)}^{[3]}=(\boldsymbol{\mathbf{I}}_{2}-\boldsymbol{\Pi})\otimes\boldsymbol{\Pi}\otimes(\boldsymbol{\mathbf{I}}_{2}-\boldsymbol{\Pi})\label{eq:OrthProj3}
\end{equation}
and for the logical projectors one has:

\begin{equation}
\boldsymbol{A}^{[3]}=\boldsymbol{\Pi}\otimes\mathbf{\mathbf{I}}_{2}\otimes\mathbf{\mathbf{I}}_{2}\qquad\boldsymbol{B}^{[3]}=\mathbf{\mathbf{I}}_{2}\otimes\boldsymbol{\Pi}\otimes\mathbf{\mathbf{I}}_{2}\qquad\boldsymbol{C}^{[3]}=\mathbf{\mathbf{I}}_{2}\otimes\mathbf{\mathbf{I}}_{2}\otimes\boldsymbol{\Pi}\label{eq:LogProj3}
\end{equation}

For arity $n=3$ the conjunction $AND$ becomes then straightforwardly:

\begin{equation}
\boldsymbol{F}_{AND}^{[3]}=\boldsymbol{A}^{[3]}\cdot\boldsymbol{B}^{[3]}\cdot\boldsymbol{C}^{[3]}=\left(\begin{array}{cccccccc}
0 & 0 & 0 & 0 & 0 & 0 & 0 & 0\\
0 & 0 & 0 & 0 & 0 & 0 & 0 & 0\\
0 & 0 & 0 & 0 & 0 & 0 & 0 & 0\\
0 & 0 & 0 & 0 & 0 & 0 & 0 & 0\\
0 & 0 & 0 & 0 & 0 & 0 & 0 & 0\\
0 & 0 & 0 & 0 & 0 & 0 & 0 & 0\\
0 & 0 & 0 & 0 & 0 & 0 & 0 & 0\\
0 & 0 & 0 & 0 & 0 & 0 & 0 & 1
\end{array}\right)\label{eq:OperAND3}
\end{equation}
$R$\\

For arity $n=3$ the majority $MAJ$ operator will be a $8\times8$
matrix, its expression can written directly using equation (\ref{eq:MAJ3})
and equation (\ref{eq:LogProj3}):

\begin{equation}
\boldsymbol{F}_{MAJ}^{[3]}=\boldsymbol{A}^{[3]}\cdot\boldsymbol{B}^{[3]}+\boldsymbol{A}^{[3]}\cdot\boldsymbol{C}^{[3]}+\boldsymbol{B}^{[3]}\cdot\boldsymbol{C}^{[3]}-2\boldsymbol{A}^{[3]}\cdot\boldsymbol{B}^{[3]}\cdot\boldsymbol{C}^{[3]}=\left(\begin{array}{cccccccc}
0 & 0 & 0 & 0 & 0 & 0 & 0 & 0\\
0 & 0 & 0 & 0 & 0 & 0 & 0 & 0\\
0 & 0 & 0 & 0 & 0 & 0 & 0 & 0\\
0 & 0 & 0 & 1 & 0 & 0 & 0 & 0\\
0 & 0 & 0 & 0 & 0 & 0 & 0 & 0\\
0 & 0 & 0 & 0 & 0 & 1 & 0 & 0\\
0 & 0 & 0 & 0 & 0 & 0 & 1 & 0\\
0 & 0 & 0 & 0 & 0 & 0 & 0 & 1
\end{array}\right)
\end{equation}

\subsection{Selection operators}

The method for selecting eigenvalues is similar to the one for elective
functions given in equation (\ref{eq:ElEqProj}). Because the projectors
of the type $\boldsymbol{\Pi}_{(a,b,c...)}^{[n]}$ are rank-$1$ projectors,
the product (matrix product) with whatever other commuting projection
operator (for example the logical operator $\boldsymbol{F}_{i}^{[n]}$)
will also give a rank-$1$ projector and more precisely this will
be the same projector multiplied by the eigenvalue. So for whatever
logical operator $\boldsymbol{F}_{i}^{[n]}$ of the considered family
one has:

\begin{equation}
\boldsymbol{F}_{i}^{[n]}\cdot\boldsymbol{\Pi}_{(a,b,c...)}^{[n]}=f_{i}^{[n]}(a,b,c,...)\:\boldsymbol{\Pi}_{(a,b,c...)}^{[n]}\label{eq:EigenvalProjSel-2}
\end{equation}
On the right of equation (\ref{eq:EigenvalProjSel-2}) the truth value
is multiplied by the corresponding rank $1$ projector.

To get explicitly the eigenvalue one can take the trace of the product
of the two operators on the left of equation (\ref{eq:EigenvalProjSel-2}).
In this way one obtains the truth value $f_{i}^{[n]}(a,b,c...)$ corresponding
to a case of a fixed combination of the values $(a,b,c,...)^{[n]}$
of the logical arguments (an \textit{interpretation}).

The method for selecting eigenvalues is similar to the one for elective
functions given in equation (\ref{eq:ElEqProj}). Because the projectors
of the type $\boldsymbol{\Pi}_{(a,b,c...)}^{[n]}$ are rank-$1$ projectors,
the product (matrix product) with whatever other projector (for example
the logical operator $\boldsymbol{F}_{i}^{[n]}$) will also give a
rank-$1$ projector and more precisely this will be the same projector
multiplied by the eigenvalue. So for whatever logical operator $\boldsymbol{F}_{i}^{[n]}$of
the considered family the truth value is multiplied by the corresponding
rank-$1$ projector.

\section{Eigenvectors, eigenvalues and truth values}

Starting with the two-dimensional rank-$1$ projector $\boldsymbol{\Pi}$
for the one-argument case, vectors $\overrightarrow{(0)}$ and $\overrightarrow{(1)}$
are $2$-dimensional orthonormal column vectors as shown in equations
(\ref{eq:eigenvector2d}).

The choice of the position of the value $1$ in the column follows
the quantum information convention for a ``qubit-$1$'' \cite{key-94}.
The Dirac bra-ket notation $|\psi>$ representing vectors used in
quantum mechanics, (\textit{i.e.} would have been here: $|0>\equiv\overrightarrow{(0)}$
and $|1>\equiv\overrightarrow{(1)}$) has not been used here purposely
in order to show that this method is not only restricted to problems
related with quantum physics.

For the two-argument case $n=2$ the vectors will have the dimension
$2^{n=2}=4$ and the complete family of $16$ commuting projection
operators represents all possible logical propositions and will be
interpretable when applied on the four possible orthonormal eigenvectors
of this family that form the complete canonical basis. These vectors
will be represented by the symbol $\overrightarrow{(a,b)}$, where
the arguments $a,b$ take the values $\{0,1\}$ and represent one
of the four possible cases:

\[
\overrightarrow{(0,0)}=\overrightarrow{(0)}\otimes\overrightarrow{(0)}=\left(\begin{array}{c}
1\\
0\\
0\\
0
\end{array}\right)\qquad\overrightarrow{(0,1)}=(0)\otimes\overrightarrow{(1)}=\left(\begin{array}{c}
0\\
1\\
0\\
0
\end{array}\right)
\]
\begin{equation}
\overrightarrow{(1,0)}=\overrightarrow{(1)}\otimes\overrightarrow{(0)}=\left(\begin{array}{c}
0\\
0\\
1\\
0
\end{array}\right)\qquad\overrightarrow{(1,1)}=\overrightarrow{(1)}\otimes\overrightarrow{(1)}=\left(\begin{array}{c}
0\\
0\\
0\\
1
\end{array}\right)
\end{equation}

When applying the logical projection operators on these vectors the
resulting eigenvalue is the truth value of the corresponding logical
proposition meaning that operations on the eigenspace of a logical
observable family are interpretable. For example for $n=2$ arguments
the complete family of $16$ commuting logical observables represents
all possible logical connectives and operations are interpretable
when applied to one of the four possible canonical eigenvectors of
the family. These vectors, corresponding to all the possible interpretations,
are represented by the vectors $\overrightarrow{(0,0)}$, $\overrightarrow{(0,1)}$,
$\overrightarrow{(1,0)}$ and $\overrightarrow{(1,1)}$ forming a
complete orthonormal basis.

\noindent Now what happens when the state-vector is not one of the
eigenvectors of the logical system? One can always express a normalized
vector as a decomposition on a complete orthonormal basis. In particular
one can express it over the canonical eigenbasis of the logical observable
family. For two-arguments this vector can be written as: 
\[
\overrightarrow{(\phi)}\,\,=\,\,C_{00}\,\overrightarrow{(0,0)}\,+\,C_{01}\,\overrightarrow{(0,1)}\,+\,C_{10}\,\overrightarrow{(1,0)}\,+\,C_{11}\,\overrightarrow{(1,1)}
\]
When only one of the coefficients is non-zero (in this case its absolute
value must take the value $1$) then one is back in the preceding
situation of a determinate interpretation (determinate input atomic
propositional case). But when more than one coefficient is non-zero
one is in a ``mixed'' or ``fuzzy'' case. Such a state can be considered
as a coherent superposition of interpretations. This can lead to a
fuzzy-logic treatment as was proposed in \cite{key-9}, fuzzy Logic
deals with truth values that may be any number between $0$ and $1$,
here the truth of a proposition may range between completely true
and completely false.

An important remark is that the choice of the eigenbasis is not fixed,
meaning that for every choice there is a complete family of logical
projection operators, so as stated above one could imagine working
with two (or more) logical systems characterized each by their family
of projective operators. The operators of one family do not (generally)
commute with the operators of another family. This non-commuting property
has its analogue in the general quantum mechanical treatment. Without
extending this argument further one sees the potentiality of considering
this kind of approach keeping in mind that in linear algebra basis
change is obtained by means of unitary operators and this is somewhat
at the heart of quantum computation where all logical operations are
done by means of unitary transformations and by measurements using
projection operators.

\section{Properties of Eigenlogic}

To summarize, all the logical projection operators have the following
properties in Eigenlogic. 
\begin{enumerate}
\item The dimension of the vector space spanned by the logical operators
is $d_{n}=2^{n}$. All logical projection operator of the same family
are $d_{n}\times d_{n}$ square matrices. 
\item All logical operators are idempotent projection operators (see eq.
(\ref{eq:IdempotProj})). This means that in the logical eigenbasis
of the family the matrices are diagonal with eigenvalues either $0$
or $1$. 
\item All the logical projection operators of a given family are commutative
pairwise. This means that all the respective matrices are diagonal
on the logical eigenbasis of the family. 
\item The logical projection operators are not necessarily orthogonal. This
means that the matrix product of two logical operators is not necessarily
the nil operator. 
\item The number of different logical projection operators of a given family
is $2^{2^{n}}$, representing a complete system of logical propositions.
This number corresponds to the number of different commuting diagonal
matrices obtained for all the combinations of $0$'s and $1$'s on
the diagonal of the matrices. 
\item For each family there are $2^{n}$ orthogonal rank-$1$ projection
operators spanning the entire vector space. The corresponding matrices
will have a single eigenvalue of value $1$, the other eigenvalues
being $0$. 
\item Every logical operator can be expressed as an elective decomposition
using the $2^{n}$ orthogonal rank-$1$ projection operators, where
the coefficients of the decomposition can only take the values $0$
or $1$ (see eq. (\ref{eq:OperatDecomp2}) for $n=2$). \cite{key-2} 
\item Every rank-$1$ projector of the family can be obtained by the means
of the Kronecker product, the seed projector $\boldsymbol{\Pi}$ and
its complement $(\mathbf{\mathbf{I}}_{2}-\boldsymbol{\Pi})$ (see
eq. (\ref{eq:projPi}), eq. (\ref{eq:OrthProj2}) and eq. (\ref{eq:OrthProj3})). 
\item The negation of a logical operator, which is its complement, is obtained
by subtracting the operator from the identity operator (see eq. (\ref{eq:OperatorNeg})). 
\item The eigenvectors of the family of $n$-arity commuting logical projection
operators form an orthonormal complete basis of dimension $d_{n}=2^{n}$.
This basis corresponds to the canonical basis and each eigenvector
corresponds to a certain combination of logical arguments, named an
\textit{interpretation}, of the logical propositional system. 
\item The eigenvalues of the logical operators are the truth values of the
respective logical proposition and each eigenvalue is associated to
a given eigenvector corresponding to an interpretation of the input
atomic proposition. 
\item The truth value of a given logical operator for a given interpretation
of $n$ arguments can be obtained using equation (\ref{eq:EigenvalProjSel-2}).\\
\end{enumerate}

\section{Discussion and related work}

Attempts to link geometry to logic are very numerous and date back
to the first efforts to formalize logic. The most celebrated ones
are for example Aristotle's square of oppositions for the 4 categorical
propositions (Subject-copula-Predicate), Leonhard Euler's (1707-1783)
diagrams illustrating propositions and quantifiers (all, no, some,\dots ),
C. L. Dodgson's (alias Lewis Carroll 1832-1898) diagrams seeking symmetry
for true and false having a striking resemblance with modern Karnaugh
maps and of course the methods developed by John Venn \cite{key-12}
which were mentioned above.

In modern logic design methods, truth tables, Karnaugh maps, hypercubes,
logic and threshold networks, decision trees and diagram graphs, are
extensively used for representing Boolean data structures \cite{key-15}.
Logical reduction based on symmetry is a very important topic which
uses Hesse diagrams, Shannon and Davio expansions and the Post theorems
on symmetries of Boolean functions. Vectorization is also a standard
procedure in logic for example using \textit{truth vectors} and \textit{carrier
vectors} (reduced truth vectors of symmetric Boolean functions).

In the following are briefly quoted recent researches which came up
during this investigation and which support the approach based on
linear algebra presented in this paper.

Starting with \textit{Matrix Logic} developed by August Stern \cite{key-22-1}
which gives directly a matrix formulation for logical operators, by
putting the truth values as matrix coefficients, in the way of Karnaugh
diagrams. So for example a two argument logical function becomes a
$2\times2$ matrix, this is a fundamental difference when compared
with the method given here above where $4\times4$ matrices are used.
Using scalar products on vectors and mean values on operators, this
formalism gives a method to resolve logical equations and allows to
enlarge the alphabet of the truth-values with negative logic antivalues.

A breakthrough has been undoubtedly made by \textit{Vector Logic}
developed by Eduardo Mizraji \cite{key-24}. This approach vectorizes
logic where the truth values map on orthonormal vectors. Technically
this approach is different from the one presented in this paper because
the resulting operators for $2$ arguments are represented by $2\times4$
matrices and do not represent projection operators. Vector logic can
also handle three-valued logic and applications have been proposed
for neural networks.

A very pertinent development, which is close to the approach in this
paper, was done by Vannet Aggarwal and Robert Caldebrabnk \cite{key-20}
in the framework of quantum error-coding theory, their work was also
justified by the \textit{Projection Logic} formulation of David Cohen
\cite{key-22}. In their method they connect Boolean logic to projection
operators derived initially from the Heisenberg-Weyl group. They associate
the dimension of the considered projector with the Hamming weight
(number of $1$'s in the truth table) of the corresponding Boolean
function. The logical operators they obtain are commuting projectors,
as in the work presented here.

The idea of linking logic and linear algebra is also becoming natural
because of the research effort due to the promise that quantum theory
can bring to fields outside of physics, principally in computer science.
Of course one must consider the quantum computer quest but also more
recent developments in other research areas such as semantic web information
retrieval \cite{key-97,key-98} and machine learning \cite{key-99}.
All these methods lie on linear algebra methods using vectors and
operators in Hilbert space.

Recently the concept of \textit{quantum predicate} introduced by E.
D'Hondt and P. Panangaden \cite{key-96} proposes an interpretation
similar to the one presented here. As stated by Mingsheng Ying in
\cite{key-95}: ``In classical logic, predicates are used to describe
properties of individuals or systems... then what is a quantum predicate?''
; ``... a quantum predicate is defined to be a physical observable
represented by a Hermitian operator with eigenvalues within the unit
interval\textquotedblright .

\section{Conclusion and perspectives}

In the formulation given here a more general method is proposed, enabling
the construction of logical projectors from a single seed projection
operator using the Kronecker product. It gives also a simpler formulation
because George Boole's elective interpretation of logic shows that
the idempotence property (\ref{eq:IndexLaw}) and (\ref{eq:idempotence})
in association with distributivity (\ref{eq:Distributiv}) and commutativity
(\ref{eq:Commutativ}) permit to identify directly commuting projection
operators with logical functions.

The formulation of logic presented here is named \textit{Eigenlogic},
it uses operators in linear algebra as propositions and is linked
to the formulation of elective symbolic algebra of George Boole in
\cite{key-1}. This similarity is striking and is more than just an
analogy, as justified here-above, at the heart of this is the idempotence
property. The logical operators belong to families of commuting projection
operators. The interesting feature is that the eigenvalues of these
operators are the truth values of the corresponding logical connective,
the associated eigenvectors corresponding to one of the fixed combination
of the inputs (interpretations). The outcome of a ``measurement''
or ``observation'' on a logical observable will give the truth value
of the associated logical proposition, and becomes ``interpretable''
when applied to its eigenspace leading to a natural analogy with the
measurement postulate in Quantum Mechanics. The following diagram
summarizes this point of view:\\

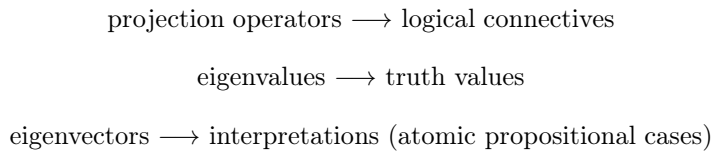
\begin{center}
projection operators $\longrightarrow$ logical connectives 
\par\end{center}

\begin{center}
eigenvalues $\longrightarrow$ truth values
\par\end{center}

\begin{center}
eigenvectors $\longrightarrow$ interpretations (atomic propositional
cases)\\
\par\end{center}

Some precision must be given concerning the last line of the diagram,
the word \textit{intepretation} is meant in the way used in logic:
an \textit{interpretation} is an assignment of truth values for each
atomic proposition that occurs in a \textit{well-formed formula}.
A \textit{well-formed formula} being a complex formula containing
exclusively logical connectives. This means that the set of atomic
propositions can have different interpretations, the ones leading
to the \textit{satisfaction} of a logical proposition (a proposition
is satisfied when it is true) are called the \textit{models }(\textit{n.b.}
sometimes the word model is used more generally as a synonymous of
the word interpretation).

A theoretical justification and a link to quantum mechanics can also
be found in Pierre Cartier \cite{key-50}, relating the link between
the algebra of logical propositions and the set of all valuations
on it, he writes: ``...in the \textit{theory of models} in logic
a model of a set of propositions has the effect of validating certain
propositions. With each logical proposition one can associate by duality
the set of all its true valuations represented by the number $1$.
This correspondence makes it possible to interpret the algebra of
propositions as a class of subsets, conjunction and disjunction becoming
respectively the intersection and union of sets. This corresponds
to the \textit{Stone duality} proved by the Stone representation theorem\textit{
}and is one of the spectacular successes of twentieth century mathematics....The
development of quantum theory led to the concept of a quantum state,
which can be understood as a new embodiment of the concept of a valuation''.
The idea is not new, as was discussed before and in \cite{key-93},
and stems from John Von Neumann's proposal of ``projections as propositions''
in \cite{key-92} which was subsequently formalized in quantum logic
with Garret Birkhoff in \cite{key-91}. 

Concerning the first line of the diagram one can generalize to eigenvalues
different from the couple $\{0,1\}$ associated to projection operators
for example by using the couple $\{+1,-1\}$ associated to self-inverse
unitary operators this has been done in \cite{key-9}, in general
one can associate a binary logical operator with whatever couple of
distinct eigenvalues $\{\lambda_{1},\lambda_{2}\}$ the corresponding
family of logical operators can be found by matrix interpolation methods
as proposed in \cite{key- 29}.

In propositional logic the arguments of a compound logical proposition
are the atomic propositions, in Eigenlogic, these are what we have
named the \textit{logical projector operators} (also sometimes named
\textit{dictators} in logic \cite{key-9,key-82}).\textit{ }Examples
are the one-argument logical projector $\boldsymbol{A}$ in equation
(\ref{eq:LogOperator1}); the two two-argument logical projectors
$\boldsymbol{A}^{[2]}$ and $\boldsymbol{B}^{[2]}$ in equations (\ref{eq: LogProj2A},\ref{eq:LogProj2B});
the three three-argument logical projectors $\boldsymbol{A}^{[3]}$,
$\boldsymbol{B}^{[3]}$ and $\boldsymbol{C}^{[3]}$ in equation (\ref{eq:LogProj3})
and so on for higher arity.

This is a fundamental difference with what is usually considered in
quantum logic (for a definition of atomic propositions in quantum
logic see \textit{e.g.} \cite{key-93} p. 98) where atomic propositions
are associated with rays \textit{i.e.} quantum pure state density
matrices. In Eigenlogic the logical connective conjunction ($AND$$,\,\wedge$),
which is non-atomic, is represented by a ray (rank-$1$ projection
operator), see equations (\ref{eq:OperAND2}, \ref{eq:OperAND3}),
the other $n-1$ rays are simply obtained by complementing selectively
the arguments of the conjunction. In general, here, rays correspond
to Kronecker products of generating projection operators (seed projection
operator), see equations (\ref{eq:OrthProj2}) and (\ref{eq:OrthProj3})
and are non-atomic (except in the case of one argument: $n=1$). From
the point of view of logic atomic propositions must be independent
propositions and this can only be achieved with the formulation given
by (\ref{eq: LogProj2A}) and (\ref{eq:LogProj2B}) and not by mutually
exclusive projection operators, such as the rank-$1$ projection operators
which are thus not independent. Thus for Eigenlogic, atomic propositions
are not rays when considering connectives with more than one argument
($n\geq2$). 

In this work complete logical families of commuting projection operators
correspond to compatible propositions this is also a difference with
quantum logic. As mentioned by David W. Cohen (p. 37 \cite{key-22})
``A quantum logic is a logic with at least two propositions that
are not compatible''. In future research the interplay of logical
observables which do not belong to the same compatible logical family
of commuting observables will be considered, this could bring insights
for quantum logic and quantum computation and address the important
topic of quantum non-contextuality.

An algorithmic approach for logical connectives with a large number
of arguments could be interesting to develop using the Eigenlogic
observables in high-dimensional vector spaces. But because the space
grows in dimension very quickly, it may not be particularly useful
for practical implementation without logical reduction. It would be
interesting to develop specific algebraic reduction methods for logical
observables inspired from actual research in the field. For a good
synthesis of the state of the art, see \textit{e.g.} \cite{key-15}.

Applications in the domain of information retrieval for applications
in semantic Web seem possible. The \textit{Quantum Interaction} community
through annual conferences promotes the links between quantum mechanics
and fields outside physics with many applications in social sciences
\cite{key-89}. The methods are based upon the exploitation of the
mathematical formalism, basely linear algebra in Hilbert space, of
quantum mechanics \cite{key-97} combined with the peculiar aspects
of the quantum postulates. Applications are found in modern semantic
theories such as distributional semantics or in connectionist models
of cognition \cite{key-88}.

More generally we think that this view of logic could add some insight
on more fundamental issues. Boolean functions are nowadays considered
as a ``toolbox'' for resolving many problems in theoretical computer
science, information theory and even fundamental mathematics. In the
same way Eigenlogic can be considered as a new ``toolbox''.

\section{Acknowledgements}

I wish to thank my colleague and friend François Dubois, Mathematician
from CNAM Paris (FR), with whom I have an ongoing collaboration on
quantum modeling and to whom is due the idea of using classical interpolation
methods for the logical expansions which heve been applied for multivalued
logic in \cite{key-9}. I am also very grateful to Francesco Galofaro,
Semiotician from Politecnico Milano (IT) and Free University of Bolzano
(IT) for his pertinent advices on semantics and logic some of the
ideas are being introduced linking quantum computing, information
retrieval and semantics. I also want to associate my colleague Bich-Lien
Doan of CentraleSupélec and LRI (Laboratoire de Recherche en Informatique)
for being at the origin of some of this multidisciplinary research.

I much appreciated the feedback from the Quantum Interaction community
for the work these last years (\cite{key-9,key-98}), the historical
aspects of the work presented here were outlined at the QI-2016 conference
in San Francisco and I wish to thank the organizers and the technical
committee particularly José Acacio da Barros of San Francisco State
University (CA, USA) and Ehtibar Dzhafarov of Purdue University (IN,
USA). In this community I appreciated the fruitful discussions with
Peter Bruza from QUT (Brisbane AUS), Emmanuel Haven and Sandro Sozzo
of University of Leicester (UK), Andrei Khrennikov of Linnaeus University
(SWE), Dominic Widdows from Microsoft Bing Bellevue (WA, USA), Trevor
Cohen from University of Texas at Houston (TX, USA), Peter Wittek
from ICFO (Barcelona ESP) and finally Keith van Rijsbergen from University
of Glasgow (UK) which was perhaps at the origin of this research because
of his comment at the conclusion of the QI-2012 conference in Paris
about the fact that George Boole already had geometrical insights
on the representation of logical functions, especially negation, in
a vector space...

I want to thank the reviewer of my first journal submission for his
deep analysis of this work and for bringing very constructive critics
to the original work of George Boole, many of his remarks have been
introduced in this version. And finally I want to thank Stanlety Burris
of the University of Waterloo (CAN) with whom I had a correspondence
and who highlighted the contribution of Theodore Halperin, many of
his remarks have been included here.


\begin{thebibliography}{10}
\bibitem{key-1} George Boole, ``The Mathematical Analysis of Logic.
Being an Essay To a Calculus of Deductive Reasoning'', (1847), (reissued
Ed. Forgotten Books ISBN 978-1444006642-9).

\bibitem{key-2} George Boole, ``An Investigation of the Laws of
Thought on Which are Founded the Mathematical Theories of Logic and
Probabilities'', Macmillan (1854) (reissued by Cambridge University
Press, 2009; ISBN 978-1-108-00153-3).

\bibitem{key-3} Maria Panteki, ``The Mathematical Background of
George Boole's Mathematical Analysis of Logic (1847)'', J. Gasser
(ed.), A Boole Anthology, 167-212, Kluwer Academic Publishers, (2000).

\bibitem{key-4} Theodore Halperin, ``Boole's Logic and Probability,
a Critical Exposition from the Standpoint of Contemporary Logic and
Probability Theory'', North Holland, (1976) II ed. (1986).

\bibitem{key-5} Theodore Halperin, ``Boole's Algebra isn't Boolean
Algebra. A Description Using Modern Algebra, of What Boole Really
Did Create'', Mathematics Magazine 54(4): 172\textendash 184 (1981).
Reprinted in A Boole Anthology ed. James Gasser. Synthese Library
volume 291, Spring- Verlag. (2000).here

\bibitem{key-6} Charles Sanders Peirce, ``On the Algebra of Logic:
A Contribution to the Philosophy of Notation\textquotedblright , American
Journal of Mathematics, Volume 7, (1885).

\bibitem{key-7} Emil Post, ``Introduction to a General theory of
Elementary Propositions \textquotedblright , American Journal of Mathematics
43: 163\textendash 185, (1921).

\bibitem{key-8} Ludwig Wittgenstein, ``Logisch-Philosophische Abhandlung\textquotedblright ,
Annalen der Naturphilosophie, Ed. Wilhelm Ostwald, Wien (1921), ``Tractatus
Logico-Philosophicus\textquotedblright , translated and published
in bilingual edition, Routledge \& Kegan Paul, London, (1922).

\bibitem{key-8-1} Karl Menger, ``Reminiscences of the Vienna Circle
and the Mathemathical Colloquium\textquotedblright{} (1942), Editors:
L. Golland B.F. McGuinness, Sklar, Ap.be - Springer (1994).

\bibitem{key-41} John Corcoran, ``Aristotle's Prior Analytics and
Boole's Laws of Thought'', History and Philosophy of Logic, 24, pp.
261\textendash 288. (2003).

\bibitem{key-9} François Dubois and Zeno Toffano, ``Eigenlogic:
a Quantum View for Multiple-Valued and Fuzzy Systems\textquotedblright ,
Quantum Interaction. QI 2016. Lecture Notes in Computer Science, vol
10106. Springer, pp. 239-251, 2017, arXiv:1607.03509 {[}quant-ph{]}.

\bibitem{key-10} Donald E. Knuth, ``The Art of Computer Programming'',
Volume 4, Fascicle 0: Introduction to Combinatorial Algorithms and
Boolean Functions, Ed. Addison-Wesley Professional, (2009).(1847)

\bibitem{key-21} Stanley Burris, (2000). ``The Laws of Boole's Thought''.
Mdictatoranuscript (http://www.math.uwaterloo.ca/\textasciitilde{}snburris/htdocs/myworks/preprints/aboole.pdf)
(2000), and private correspondence.

\bibitem{key-11} Edward V. Huntington, ``Sets of independent postulates
for the algebra of logic''. Trans. AMS 5:288\textendash 309 (1904).

\bibitem{key-12} John Venn, ``Symbolic Logic'', London: Macmillan
and Company, ISBE. D'Hondt and P. Panangaden 1-4212-6044-1. (1881).

\bibitem{key-31}Schroder, E., ``Vorlesungen über die Algebra der
Logik'', Vol. I, Anh. 6, B.G. Teubner, Leipzig. (1890)

\bibitem{key-13} Hassler Whitney, ``Characteristic functions and
the algebra of logic'' in Annals of Mathematics 34 (1933), pp. 40-414

\bibitem{key-15} Svetlana N. Yanushkevich, Shmerko, V.P.: ``Introduction
to Logic Design''. CRC Press (2008)

\bibitem{key-14} Howard H. Aiken, ``Synthesis of electronic computing
and control circuits\textquotedblright{} Ann. Computation Laboratory
of Harvard University, XXVII, Harvard University, Cambridge, MA, (1951).

\bibitem{key-51} Shannon, C. E. \textquotedbl{}A Symbolic Analysis
of Relay and Switching Circuits\textquotedbl{}. Trans. AIEE. 57 (12):
713\textendash 723. (1938)

\bibitem{key-61} Marshall H. Stone, ``Linear Transformations in
Hilbert Space and Their Applications to Analysis\textquotedblright ,
p.70: ``Projections\textquotedblright . (1932)

\bibitem{key-71} Marshall H. Stone. ``The theory of representation
for Boolean algebras''. Transactions of the American Mathematical
Sdictatorociety, 40(1):37\textendash 111, Jul. (1936)

\bibitem{key-72} Marshall H. Stone. ``Applications of the theory
of Boolean rings to general topology''. Transactions of the American
Mathematical Society, 41(3):375\textendash 481, May (1937)

\bibitem{key-81} Dirk Schlimm, ``Bridging Theories with Axioms:
Boole, Stone, and Tarski'', New Perspectives on Mathematical Practices,
World Scientific pp. 222-235, (2009)

\bibitem{key-91} Garret Birkhoff, John von Neumann: ``The Logic
of Quantum Mechanics\textquotedblright . The Annals of Mathematics,
2nd Ser., 37 (4), 823-843 (1936)

\bibitem{key-92} John von Neumann, ``Mathematische Grundlagen der
Quantenmechanik. Grundlehren der mathematischen Wissenschaften'',
volume Bd. 38. (Springer, Berlin, 1932) 106. ``Mathematical Foundations
of Quantum Mechanics''. Investigations in Physics, vol. 2. (Princeton
University Press, Princeton, 1955)

\bibitem{key-93} François David, ``The Formalisms of Quantum Mechanics,
An Introduction'', Springer Lecture Notes in Physics, ISBN 978-3-319-10538-3,
(2015)

\bibitem{key-94} Nielsen, M.A., Chuang, I.L.: Quantum Computation
and Quantum Information. Cambridge University Press (2000)

\bibitem{key-22-1}August Stern, ``Matrix logic'', North-Holland,
(1988).

\bibitem{key-24} Eduardo Mizraji, ``Vector logics: the matrix-vector
representation of logical calculus''. Fuzzy Sets and Systems, 50,
179\textendash 185, (1992).

\bibitem{key-20} Vaneet Aggarwal and Robert Calderbank, ``Boolean
functions, projection operators, and quantum error correcting codes,\textquotedblright{}
in Proc. Int. Symp. Inf. Theory, Nice, France, pp. 2091\textendash 2095,
(2007).

\bibitem{key-82} Ryan O'Donnell, ``Analysis of Boolean Functions'',
Cambridge University Press, 2014.

\bibitem{key-22} David W. Cohen, ``An introduction to Hilbert space
and quantum logic,\textquotedblright{} Springer-Verlag, (1989).

\bibitem{key-97} Keith van Rijsbergen, ``The Geometry of Information
Retrieval'', Cambridge University Press, Cambridge (2004)

\bibitem{key-98} Barros, J., Toffano, Z., Meguebli, Y., Doan, B.-L.,
``Contextual query using bell tests'', In: Atmanspacher, H., Haven,
E., Kitto, K., Raine, D. (eds.) QI 2013. LNCS, vol. 8369, pp. 110\textendash 121.
Springer, Heidelberg (2014). doi:10.1007/ 978-3-642-54943-4 10

\bibitem{key-99} Peter Wittek, ``Quantum Machine Learning. What
Quantum Computing Means to Data Mining'', Academic Press Elsevier,
Amsterdam, 2014

\bibitem{key-96} E. D'Hondt and P. Panangaden, ``Quantum weakest
preconditions''. Mathematical Structures in Computer Science, 16,
pp. 429-451,(2006).

\bibitem{key-95} Ying, M.S., ``Foundations of Quantum Programming'',
Morgan Kaufmann, (2016).

\bibitem{key-50} Pierre Cartier, ``A mad day's work: from Grothendieck
to Connes and Kontsevich The evolution of concepts of space and symmetry\textquotedblright ,
Journal: Bull. Amer. Math. Soc. 38 (2001), 389-408.

\bibitem{key- 29} Zeno Toffano and François Dubois: ``Interpolation
Methods for Binary and Multivalued Logical Quantum Gate Synthesis'',
presented at TQC2017, Paris, France, June 14-16, 2017, arXiv:1703.10788
{[}quant-ph{]}

\bibitem{key-89} Emmanuel Haven, Andrei Khrennikov, ``Quantum Social
Science'', Cambridge University Press, (2013)

\bibitem{key-88} Busemeyer, J.R., Bruza, P.D., ``Quantum models
of cognition and decision'', Cambridge University Press (2012)
\end{thebibliography}
\end{document}